\pdfoutput=1
\documentclass[11pt]{article}

\usepackage[final]{acl}

\usepackage{times}
\usepackage{latexsym}

\usepackage[T1]{fontenc}
\usepackage[utf8]{inputenc}
\usepackage{microtype}
\usepackage{inconsolata}
\usepackage{fontawesome5}
\usepackage{simpleicons}
\usepackage{graphicx}
\usepackage{booktabs}
\usepackage{multicol}
\usepackage{multirow}
\usepackage{graphicx}
\usepackage{subfigure}
\usepackage{makecell}
\usepackage{amsmath,mathtools,amsthm}
\usepackage{color}
\usepackage{algorithm}
\usepackage{algorithmic}
\usepackage{amsfonts}
\usepackage{amsmath}
\usepackage{enumitem}
\usepackage{colortbl}
\usepackage[normalem]{ulem}  % 加载 ulem 宏包
\usepackage{pifont}
\usepackage{enumitem}
\newcommand{\ul}[1]{\underline{#1}}
\usepackage{float}

\usepackage{tcolorbox}
\tcbset{
  aibox/.style={
    width=\linewidth,
    top=10pt,
    colback=gray!13,
    colframe=black,
    colbacktitle=black,
    coltitle=white,
    center,
    title={#1},  % 用#1接收标题参数
  }
}
\newtcolorbox{AIbox}[2][]{aibox={#2},#1} 

\title{VAPO: End-to-end Slide-Enhanced Speech Recognition with Omni-modal Large Language Models}
\author{
Rui Hu\textsuperscript{1,2},
Delai Qiu\textsuperscript{2},
Yining Wang\textsuperscript{2},
Shengping Liu\textsuperscript{2}, 
Jitao Sang\textsuperscript{1,3\thanks{Corresponding author.}} \\
\textsuperscript{1}Beijing Key Laboratory of Traffic Data Mining and Embodied Intelligence, \\Beijing Jiaotong University \\ 
\textsuperscript{2}Unisound AI Technology Co., Ltd. \\
\textsuperscript{3}State Key Laboratory of AI Safety, Beijing\\
\faEnvelope\ \texttt{\{rui.hu, jtsang\}@bjtu.edu.cn}\\
\faGithub\ \url{https://github.com/isruihu/SlideASR-Bench} \\
}
\begin{document}
\maketitle

\begin{abstract}
Omni-modal large language models (OLLMs) offer a promising end-to-end solution for slide-enhanced speech recognition due to their inherent multimodal capabilities. However, we found a fundamental issue faced by OLLMs: \textit{Visual Interference}, where models show a bias towards visible text over auditory signals, causing them to hallucinate slide content that was never spoken. To address this, we propose Visually-Anchored Policy Optimization (\textbf{VAPO}), which aims to reshape models' inference process to follow the human-like ``\textit{Look-then-Listen}'' inference chain. Specifically, we design a temporally decoupled policy: the model first extracts visual priors in a \textit{<think>} block to serve as semantic anchors, then generates the transcription in an \textit{<answer>} block. The policy  is optimized via multi-objective reinforcement learning. Furthermore, we introduce SlideASR-Bench, a comprehensive benchmark designed to address the scarcity of entity-rich data, comprising a large-scale synthetic corpus for training and a challenging real-world test set for evaluation. We conduct extensive evaluations demonstrating that VAPO effectively eliminates visual interference and achieves state-of-the-art performance on SlideASR-Bench and public datasets, significantly reducing entity recognition errors in specialized domains.

\end{abstract}

\section{Introduction}

\begin{figure}[!tbh]
    \centering
    \includegraphics[width=1\linewidth]{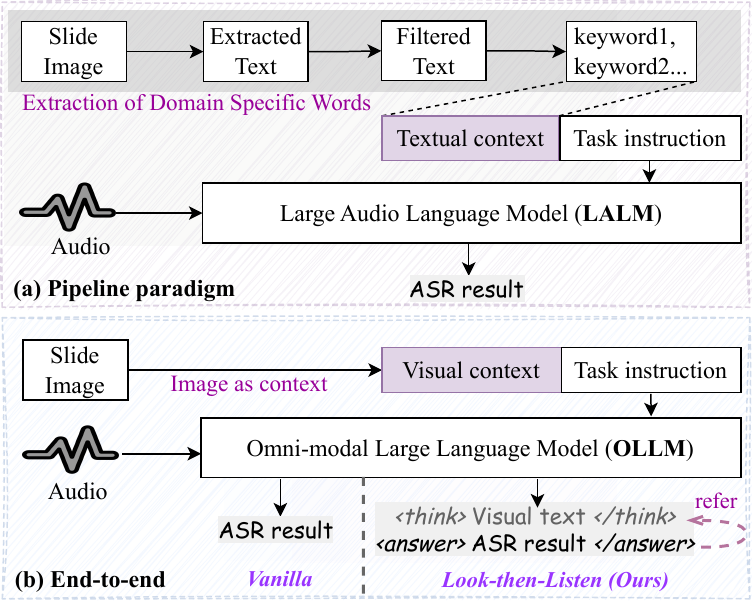}
    % \caption{Comparison of two paradigms for SlideASR task. \textbf{Top:} The pipeline paradigm, which extracts and filters domain-specific words from the image frame to serve as textual context that is then input to a LALM for ASR. \textbf{Bottom:} The end-to-end paradigm, where an OLLM directly utilizes the raw image frame as visual context along with the audio signal.}
    \caption{Comparison of paradigms for the SlideASR task. \textbf{(a) Pipeline paradigm:} Cascades independent modules which is complex. \textbf{(b) End-to-end paradigm:} Illustrates two approaches using an OLLM. The \textbf{\textit{Vanilla}} path directly generates the ASR result, often leading to visual interference. In contrast, our \textbf{\textit{Look-then-Listen}} path enforces a structured inference chain, explicitly decoupling visual perception from auditory processing.}
    \label{fig:paradigm}
\end{figure}

% 当前ASR模型对专业实体转录准确性差,
Current Automatic Speech Recognition (ASR) models, such as Whisper~\cite{radford2023whisper}, have demonstrated impressive performance in general domains. However, recognition accuracy often deteriorates significantly in specialized scenarios, such as academic lectures or technical presentations where domain-specific terminologies and rare entities are prevalent~\cite{sinhamahapatra2025slides}. 
% Lipreading方法无法利用Slide信息
While previous works have improved ASR accuracy by incorporating lip movement information~\cite{deep-avsr, conformers-avsr, AutoAVSR, AVHubert}, they ignore the rich semantic context present in presentation slides. In these scenarios, critical keywords and proper nouns are often explicitly displayed on the slides, serving as strong visual cues for the spoken content~\cite{slideavsr, slidespeech}.  For clarity, we refer to the task of improving ASR accuracy by incorporating visual context from presentation slides as \textit{SlideASR}.

% 动机段落, 介绍pipline范式及其缺点
Currently, the dominant strategy for the SlideASR task is the pipeline paradigm~\cite{slidespeech}. As  shown in Fig.~\ref{fig:paradigm}(a),  this paradigm extracts text from slide images and selects domain-specific words, which are then fed as textual context into Large Audio Language Models (LALMs)~\cite{qwen2audio, SeedASR, MiDashengLM} for ASR. This cascading process involves multiple modules, suffering from complexity and error accumulation. 

% 提出目标
Given these limitations, a natural question arises: \textbf{Can we establish an end-to-end paradigm} that directly utilizes the slide image as visual context, thus bypassing the complex pipeline?
% 引出OLLM
To answer this question, we argue that the recently emerging Omni-modal Large Language Models (OLLMs)~\cite{qwen25omni, qwen3omni, minicpmo, Megrez} offer a promising solution. OLLMs are capable of simultaneously processing textual, visual, and auditory modalities. Thus, they are inherently well-suited to accomplish the SlideASR task in an end-to-end manner.

% 指出原生OLLM在SlideASR任务上的问题
However, our preliminary investigation reveals a critical gap between this theoretical potential and practical performance (Sec.~\ref{sec:failure}). Specifically, we observe that current OLLMs suffer from severe \textbf{visual interference}. Instead of using the slide image as auxiliary information, the models exhibit a strong tendency to hallucinate, i.e., incorrectly transcribe visible slide text while this text is absent from the spoken audio. 
% 衔接句, 人类是如何听演讲的
We attribute this failure to the \textbf{lack of an intermediate reasoning phase.} Considering human behavior, when listening to professional presentations, humans typically adopt a ``\textit{Look-then-Listen}'' strategy, which temporally decouples the processing of the two modalities. We first scan the slide to establish a contextual prior of the topic and then anchor the auditory input within this context.  In contrast, vanilla OLLMs process visual and auditory signals simultaneously (left path in Fig.~\ref{fig:paradigm}(b)). Lacking this sequential guidance, strong visual cues tend to suppress auditory inputs, resulting in the observed interference.

% 方法
Inspired by this, we propose \textbf{V}isually-\textbf{A}nchored  \textbf{P}olicy  \textbf{O}ptimization (\textbf{VAPO}) which centers on reshaping the model's inference process to align with the human-like ``\textit{Look-then-Listen}'' workflow (right path in Fig.~\ref{fig:paradigm}(b)). 
Specifically, we design a \textit{<think><answer>} format to explicitly decompose the multimodal task. In the \textit{<think>} phase, the model is required to first perform Optical Character Recognition (OCR) to establish a visual context prior. Subsequently, in the \textit{<answer>} phase, it generates the transcription by attending to the audio while referencing the extracted content as a reliable anchor.
% 奖励函数
To instantiate this policy, we employ four complementary rewards: a) \textit{Format Reward} to ensure structural compliance; b) \textit{OCR Reward} to promote precise visual perception; c) \textit{ASR Reward} to maintain the overall transcription accuracy; and d) \textit{Visual Anchoring Reward} to encourage the model to effectively leverage key entities identified within the \textit{<think>} phase.

% 提出数据集
Furthermore, to address data bottlenecks, we construct \textbf{\textit{SlideASR-Bench}}. Existing datasets, e.g., SlideSpeech~\cite{slidespeech}, primarily focus on general scenarios and lack sufficient entity density. SlideASR-Bench is specifically tailored to promote the utilization of visual cues for specialized terms and comprises two subsets: 1) \textbf{\textit{SlideASR-S}}, a large-scale synthetic corpus derived from ContextASR-Bench~\cite{ContextASR-Bench} providing both training and test sets; and 2) \textbf{\textit{SlideASR-R}}, a small real-world test set for evaluation in complex presentation environments.

% 实验和贡献
We conduct extensive experiments on SlideSpeech~\cite{slidespeech} and our proposed SlideASR-Bench. Empirical results demonstrate that our approach significantly outperforms state-of-the-art models, e.g., Qwen3-Omni~\cite{qwen3omni}, particularly on entity-related metrics. The main contributions are summarized as follows: 

\begin{itemize}[leftmargin=*, ] 
    \item \textbf{Analysis:} We identify visual interference as the primary bottleneck in current OLLMs for SlideASR, revealing why naive end-to-end approaches fail. 
    \item \textbf{Method:} We propose VAPO, which reshapes the inference process into a structured, human-like ``\textit{Look-then-Listen}'' workflow, explicitly temporally decoupling visual perception from auditory transcription to eliminate visual interference.
    \item \textbf{Data:} We construct SlideASR-Bench, comprising synthetic and real-world subsets, to train and evaluate SlideASR task in entity-rich scenarios. 
\end{itemize}

\section{Visual Interference of OLLMs in End-to-end SlideASR}
\label{sec:failure}

\begin{figure}[!t]
    \centering
    \includegraphics[width=1\linewidth]{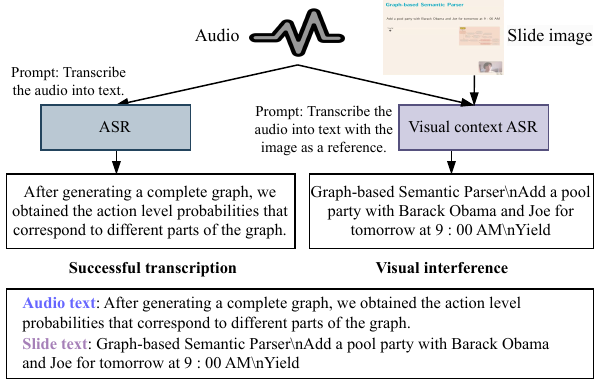}
    \caption{An illustrative example of \textbf{Visual Interference} in the end-to-end SlideASR task. \textbf{Left:} the model correctly transcribes the speech in audio-only mode. \textbf{Right:} the model erroneously outputs the text on the slide when the image is introduced as visual context.}
    \label{fig:asr2ocr}
\end{figure}

Although OLLMs theoretically have the capability to process auditory and visual modalities simultaneously, our preliminary investigation reveals a critical failure mode in their practical application. We present a motivating example in Fig.~\ref{fig:asr2ocr} using Qwen2.5-Omni-7B~\cite{qwen25omni}. In the audio-only setting, the model accurately transcribes the speech. However, when the corresponding slide image is introduced, the model's behavior shifts drastically: instead of utilizing the visual information as auxiliary context, it ignores the auditory signal and directly outputs the text visible on the slide. We term this specific failure mode as \textbf{Visual Interference}. Formally, it is defined as a phenomenon where the model generates text present in the visual context while it is absent from the speech.

To systematically quantify this phenomenon, we introduce a metric termed the \textbf{Visual Interference Rate (VIR)} which measures the tendency of the model to hallucinate text from the slide that was never spoken. We evaluate four representative OLLMs: Qwen2.5-Omni (7B/3B)~\cite{qwen25omni}, MiniCPM-o-2.6~\cite{minicpmo}, and Megrez-Omni~\cite{Megrez} on the SlideSpeech~\cite{slidespeech} dataset. The calculation follows a set-difference logic:

\begin{itemize}[leftmargin=*, noitemsep]
    \item \textbf{Step 1: Context Isolation.} Let $V_{slide}$ be the set of words on the slide and $V_{audio}$ be the set of words in the ground-truth transcript. We first identify the \textit{slide-exclusive vocabulary} $V_{exclusive} = V_{slide} \setminus V_{audio}$. These are words that appear visually but are not spoken.
    \item \textbf{Step 2: Interference Detection.} Let $V_{pred}$ be the set of words in the model's prediction. We calculate the intersection $I = V_{pred} \cap V_{exclusive}$.
    \item \textbf{Step 3: Metric Calculation.} If $I \neq \emptyset$, the sample is flagged as exhibiting visual interference. The VIR is defined as the percentage of such flagged samples in the dataset.
\end{itemize}

The quantitative results are presented in Table~\ref{tab:ocr-ratio}. All evaluated OLLMs exhibit a high VIR, ranging from 12.87\% to as high as 63.28\% on the test set. Even the strongest baseline, Qwen2.5-Omni-7B, fails to suppress visual interference in over 12\% of the samples. This consistent failure across different models indicates that this is a widespread issue rather than an isolated case. This phenomenon highlights a fundamental limitation in current paradigms: OLLMs lack an explicit mechanism to temporally decouple visual perception from auditory processing.

\begin{table}[!t]\small
\caption{Comparison of Visual Interference Rate across different OLLMs on the SlideSpeech dataset.}
    \centering
    % \resizebox{\linewidth}{!}{
    \begin{tabular}{l|ll}
    \toprule
         \multirow{2}{*}{\textbf{Model}}&   \textbf{Dev set}& \textbf{Test set}\\
                                    & \textbf{Num=1,801}&\textbf{Num=3,053}\\ \midrule
         MiniCPM-o-2.6 & 57.96\%&63.28\%\\
         Megrez-Omni& 45.14\%&44.90\%\\
         Qwen2.5-Omni-3B& 15.43\%&16.54\%\\
         Qwen2.5-Omni-7B& 13.71\%&12.87\%\\
    \bottomrule
    \end{tabular}
    % }
    \label{tab:ocr-ratio}
\end{table}

\section{Method}

\begin{figure*}
    \centering
    \includegraphics[width=1\linewidth]{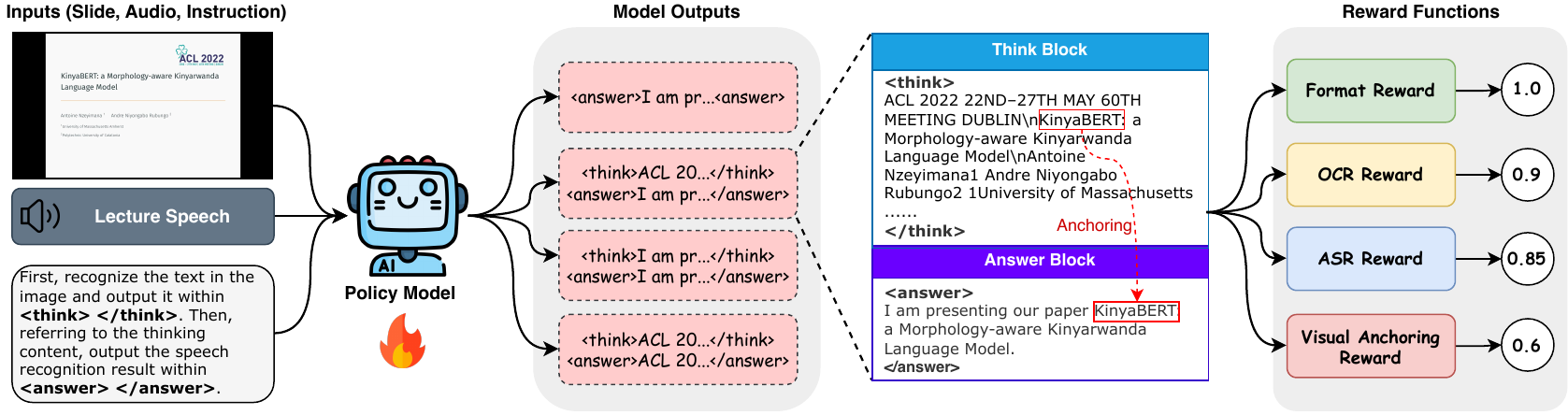}
    \caption{Overview of the Visually-Anchored Policy Optimization (VAPO) framework. The model takes audio, slide, and instruction as inputs and generates a structured \textit{<think><answer>} sequence. In the \textit{<think>} phase, the model extracts visual context, which serves as a semantic anchor (indicated by the red box) to guide the transcription in the \textit{<answer>} phase. The policy is optimized via four rewards (Format, OCR, ASR, and Visual Anchoring).}
    \label{fig:method}
\end{figure*}

To mitigate the \textit{Visual Interference} issue identified in Sec.~\ref{sec:failure}, we propose Visually-Anchored Policy Optimization (VAPO). Inspired by the human perception process, VAPO addresses this limitation by reshaping the model's inference process to achieve a temporal decoupling of modalities. Specifically, it establishes a structured \textit{<think><answer>} inference chain: the model first establishes a visual prior in the \textit{<think>} phase, and then references this content as an anchor to guide speech transcription in the \textit{<answer>} phase. This mechanism mimics the human-like ``\textit{Look-then-Listen}'' workflow. The overall framework is illustrated in Fig.~\ref{fig:method}.

\subsection{The ``Look-then-Listen'' Inference Chain}

We design a structured output format to explicitly enforce the necessary temporal decoupling between visual perception and auditory processing. Drawing inspiration from Chain-of-Thought (CoT) reasoning~\cite{o1, audioCoT}, the model is mandated to generate its output within a sequential \textit{<think><answer>} structure.

The \textit{<think>} block corresponds to the ``Look'' phase: the model first processes the visual input and is tasked with extracting textual information from the slide image. This operation establishes a critical visual context prior before the auditory processing begins. Subsequently, the \textit{<answer>} block corresponds to the ``Listen'' phase: the model generates the final transcription. In this phase, the model is required to reference the content anchored in the \textit{<think>} block. This mechanism allows specialized terms from the slide to serve as semantic anchors, thereby assisting the model in resolving ambiguous audio and enhancing overall transcription accuracy.

\subsection{Multi-Objective Policy Optimization}

To make the model follow the proposed inference chain, we design four complementary reward functions to guide the model's learning. We employ the Group Relative Policy Optimization (GRPO) algorithm~\cite{deepseekmath} for model training.

~\\
\textbf{Format Reward.} This reward aims to ensure that the model's output strictly adheres to the \textit{<think><answer>} format. A positive reward is assigned only if the model generates the complete structural tags. The reward function is defined as:
\begin{equation}
R_{\text{Format}} = 
\begin{cases} 
1, & \text{if the format is correct,} \\
0, & \text{otherwise.}
\end{cases}
\end{equation}

% ~\\
\noindent\textbf{OCR Reward.} This reward evaluates the visual perception quality in the ``Look'' phase by comparing the text generated in the \textit{<think>} block ($T_t$) with the ground truth slide text ($T_s$). We use the Word Error Rate (WER) as the metric, treating each Chinese character as a word where applicable. The reward is normalized and clipped to ensure non-negativity:
\begin{equation}
R_{\text{OCR}} = \max(1 - \text{WER}(T_t, T_s), 0).
\end{equation}

% ~\\
\noindent\textbf{ASR Reward.} This reward assesses the transcription quality in the ``Listen'' phase by comparing the output in the \textit{<answer>} block ($T_a$) with the ground truth speech transcription ($T_g$). Similarly, it is derived from the WER:
\begin{equation}
R_{\text{ASR}} = \max(1 - \text{WER}(T_a, T_g), 0).
\end{equation}

% ~\\
\noindent\textbf{Visual Anchoring Reward.} This reward bridges the ``Look'' and ``Listen'' phases by incentivizing the model to ground spoken entities in the visual context. Let $E_{target}$ be the set of critical entities present in both the ground truth slide and the ground truth speech transcript. Let $E_{gen}$ be the subset of $E_{target}$ that successfully appears in both the generated \textit{<think>} and \textit{<answer>} blocks. The reward is defined as recall of these target anchors:
\begin{equation}
R_{\text{VA}} = 
\begin{cases} 
\frac{|E_{gen}|}{|E_{target}|}, & \text{if } |E_{target}| > 0, \\
0, & \text{otherwise.}
\end{cases}
\end{equation}
This formulation encourages the model to explicitly capture and utilize the visual cues that are relevant to the current speech.

~\\
Finally, the total reward is defined as a weighted sum of the four components, where $\lambda_i$ denotes the weighting hyperparameter for each reward:
\begin{equation}
\begin{split}
R_{\text{total}} = \;& \lambda_1 R_{\text{Format}} + \lambda_2 R_{\text{OCR}} \\
& + \lambda_3 R_{\text{ASR}} + \lambda_4 R_{\text{VA}}.
\end{split}
\end{equation}

\section{SlideASR-Bench: A Benchmark for Entity-rich SlideASR Task}
\label{sec:SlideASR-Bench}

\begin{table*}[!tb] \small
    \caption{Results on the SlideSpeech, a real-world English SlideASR dataset. \textsuperscript{$\dagger$} represents results from the original paper. The best and second-best results are in \textbf{bold} and \ul{underlined}, respectively.}
    
    \centering
    \resizebox{\textwidth}{!}{
    \begin{tabular}{lcccccccc}
        \toprule
        \multirow{2}{*}{\textbf{Model}} &
        \multicolumn{4}{c}{\textbf{Dev set}} &  \multicolumn{4}{c}{\textbf{Test set}}\\
        \cmidrule(lr){2-5} \cmidrule(lr){6-9}
        & \fontsize{8}{9}{\textbf{WER}}$\downarrow$& \fontsize{8}{9}{\textbf{B-WER}}$\downarrow$& \fontsize{8}{9}{\textbf{U-WER}}$\downarrow$& \fontsize{8}{9}{\textbf{Recall}}$\uparrow$& \fontsize{8}{9}{\textbf{WER}}$\downarrow$& \fontsize{8}{9}{\textbf{B-WER}}$\downarrow$&\fontsize{8}{9}{\textbf{U-WER}}$\downarrow$& \fontsize{8}{9}{\textbf{Recall}}$\uparrow$\\\midrule
       \rowcolor{gray!13}\multicolumn{9}{c}{\textit{\textbf{Contextless}}}\\
        Qwen2-Audio& 12.56 & 12.85 & 8.72 & 91.43 & 13.19 & 13.59 &7.53 &92.91 \\
        MiniCPM-o-2.6& 16.09 & 16.68 & 8.14 & 91.98 & 18.71 & 19.41 & 8.90 &91.50 \\
        Qwen2.5-Omni-3B& 15.53 & 16.22 & 6.30 & 93.76 & 12.00 & 12.45 &5.72 &94.41 \\
        Qwen2.5-Omni-7B& 11.75 & 12.20 & 5.39 & 94.78 & 11.75 & 12.20 & 5.39 &94.78 \\
        Qwen3-Omni-30B-A3B& 10.87 & 11.31 & 5.02 & 95.04 & 11.71 & 12.21 & 4.64 &95.50 \\\midrule
        
        \rowcolor{gray!13}\multicolumn{9}{c}{\textit{\textbf{Slide text as context (Pipeline) }}}\\
        Qwen2-Audio & 139.81 & 145.05 & 69.94 & 85.40 & 146.08 & 152.41 &56.99 &88.98 \\
        Mi-Dasheng & 33.67 & 35.18 & 13.56 & 93.02 & 47.21 & 49.34 & 17.21 &91.00 \\
        Qwen3-Omni-30B-A3B& 50.43 & 52.85 & 18.05 & 96.45 & 57.12 & 59.27 & 26.75 &96.34 \\
        LCB-net\textsuperscript{$\dagger$}& 18.80 & 18.11 & 27.90 & 72.09 & 19.21 & 18.89 & 23.70 &76.48 \\
        MaLa-ASR\textsuperscript{$\dagger$}& 11.14 & 11.36 & 8.92 & 91.44 & 11.26 & 11.52 & 7.67 &92.50 \\\midrule
        
       \rowcolor{gray!13}\multicolumn{9}{c}{\textit{\textbf{Slide image as context (End-to-End)}}}\\
        MiniCPM-o-2.6& 182.96 & 192.83 & 51.07 & 86.26 & 210.37 & 220.96 &60.92 &83.22 \\
        Qwen2.5-Omni-3B& 12.22 & 12.74 & 5.26 & 95.17 & 19.99 & 20.71 & 9.80 &94.44 \\
        Qwen2.5-Omni-7B& 13.65 & 14.13 & 7.19 & 92.84 & 14.97 & 15.58 & 6.33 &93.99 \\
        Qwen3-Omni-30B-A3B& 19.85 & 20.64 & 9.30 & 95.59 & 24.13 & 24.88 & 13.44 &94.74 \\
        
        \textbf{VAPO-3B (Ours)}& \ul{9.84}& \ul{10.31}& \ul{3.61}& \ul{96.54}& \ul{10.73}& \ul{11.24}& \ul{3.55}&\ul{96.57}\\
        \textbf{VAPO-7B (Ours)}& \bf8.62& \bf9.08& \bf2.48&\bf97.61& \bf10.31& \bf10.84&\bf2.87&\bf97.32\\
    
    \bottomrule
    \end{tabular}}

    \label{tab:main-SlideSpeech}
\end{table*}

Our primary goal is to enhance ASR accuracy for domain-specific entities within visual presentation contexts. While existing datasets like SlideSpeech~\cite{slidespeech} and ChineseLips~\cite{chineselips}, provide valuable benchmarks for general-domain scenarios, we observe that they often lack a sufficient density of domain-specific named entities.  This scarcity creates a significant bottleneck for both training visually-anchored models and evaluating their capabilities in specialized scenarios. To bridge this gap, we introduce SlideASR-Bench, which comprises two distinct subsets: SlideASR-S, a large-scale synthetic corpus for training and evaluation, and SlideASR-R, a small real-world test set for stress-test.  Detailed statistics are presented in Table~\ref{tab:SlideASR-Bench-Detail}. 

~\\
\textbf{SlideASR-S.} To train and evaluate models for entity-rich scenarios, we constructed SlideASR-S by extending the ContextASR-Bench~\cite{ContextASR-Bench} dataset.  ContextASR-Bench leverages LLMs, such as DeepSeek-R1~\cite{deepseekr1}, to generate colloquial text rich in named entities based on seed text. The seed text is sourced from Named Entity Recognition (NER) datasets across multiple domains (e.g., medicine, culture, and ecology).  Text-to-speech models~\cite{CosyVoice2} are then used to convert the generated text into natural and fluent speech. 

We extract metadata for each sample from ContextASR-Bench, including the domain label $L_{\text{domain}}$ and the list of domain-specific entities $E$. Using the $L_{\text{domain}}$ and $E$ as input, we employ an LLM (e.g., Qwen2.5-14B-Instruct\footnote{https://huggingface.co/Qwen/Qwen2.5-14B-Instruct}~\cite{qwen2}) to generate a short non-colloquial, formal text in slide style. The prompt guides the LLM to include a title and key points, ensuring that all entities from the original audio are naturally embedded in the generated text. Finally, we utilize Python's Matplotlib library to render the generated text into slide images, generating a total of 8,467 samples (6,413 for training set, 2,054 for test set). We provide the prompt and data example in Appendix ~\ref{sec:app-prompts-for-slideasr-s}.

~\\
\textbf{SlideASR-R.} Furthermore, to assess the model's generalization ability in real, complex environments, we manually constructed a small-scale, high-quality, and challenging test set. We collected 60 real presentation audio clips and corresponding slide images from publicly available academic report videos, covering four specialized domains: chemistry, medicine, biology, and artificial intelligence. For each sample, we manually annotated the data by carefully comparing the speech and slide image, identifying the domain-specific entities that appear in both. We named this dataset SlideASR-R, which contains 200 domain-specific entities from real-world scenarios.

\section{Experiment}
\subsection{Experimental Setup}

\begin{table*}[!tbh]\small
    \caption{Results on the SlideASR-Bench. The best and second-best results are in \textbf{bold} and \ul{underlined}, respectively.} 
    \centering
    \resizebox{\textwidth}{!}{
    \begin{tabular}{l cccccccc}
        \toprule
        \multirow{2}{*}{\textbf{Model}} &
        \multicolumn{3}{c}{\textbf{SlideASR-S (en)}}&  \multicolumn{3}{c}{\textbf{SlideASR-S (zh)}}& \multicolumn{2}{c}{\textbf{SlideASR-R}}\\
        \cmidrule(lr){2-4} \cmidrule(lr){5-7} \cmidrule(lr){8-9}& \fontsize{8}{9}{\textbf{WER}}$\downarrow$& \fontsize{8}{9}{\textbf{NE-WER}} $\downarrow$& \fontsize{8}{9}{\textbf{NE-FNR}}$\downarrow$& \fontsize{8}{9}{\textbf{WER}}$\downarrow$& \fontsize{8}{9}{\textbf{NE-WER}}$\downarrow$& \fontsize{8}{9}{\textbf{NE-FNR}}$\downarrow$& \fontsize{8}{9}{\textbf{NE-WER}}$\downarrow$& \fontsize{8}{9}{\textbf{NE-FNR}}$\downarrow$\\ \midrule
        
        \rowcolor{gray!13}\multicolumn{9}{c}{\textit{\textbf{Contextless}}}\\ 
        Qwen2-Audio & 11.90 & 36.29 & 47.84 & 6.02 & 22.83 &40.36 & 74.56 &76.73 \\
        MiniCPM-o-2.6& 11.19 & 27.51 & 30.93 & 10.35 & 25.00 & 41.62 & 55.85 &65.37 \\
        Qwen2.5-Omni-3B& 8.37 & 24.15 & 31.04 & 4.47 & 19.89 & 38.08 & 61.31 &66.83 \\
        Qwen2.5-Omni-7B& 8.15 & 23.44 & 27.77 & 4.34 & 17.54 &32.80  & 53.68 &63.37 \\
        Qwen3-Omni-30B-A3B& 9.06 & 14.61 & 15.53 & 20.77 & 23.31 & 22.49 & 40.43 &41.09 \\ \midrule

        \rowcolor{gray!13}\multicolumn{9}{c}{\textit{\textbf{Slide text as context (Pipeline)}}}\\ 
        Qwen2-Audio & 92.16 & 66.38 & 24.82 & 39.09 & 50.58 &31.52 & 59.04 &21.29 \\
        Mi-Dasheng & 78.98 & 49.85 & 30.58 & 66.88 & 56.30 & 32.30 & 47.52 & 26.73 \\
        Qwen3-Omni-30B-A3B& 34.65 & 32.35 & 8.56 & 9.76 & 15.85 & 13.54 & 34.01 &28.22 \\ \midrule
        
        \rowcolor{gray!13}\multicolumn{9}{c}{{\textbf{Slide image as context (End-to-End)}}}\\
        MiniCPM-o-2.6& 112.90 & 49.65 & 15.01 & 89.53 & 61.25 & 45.67 & 63.73 &66.83 \\ 
        Qwen2.5-Omni-3B& 100.08 & 53.19 & 18.72 & 86.86 & 65.62 &9.62 & 49.00 &53.47 \\
        Qwen2.5-Omni-7B& 57.21 & 35.76 & 15.04 & 91.83 & 54.04 & 3.36 & 41.77 &35.15 \\
        Qwen3-Omni-30B-A3B& 101.45 & 59.64 & 12.08 & 79.21 & 46.45 & 5.54 & 32.26 &24.75 \\
        \textbf{VAPO-3B (Ours)}& \ul{4.90}& \ul{3.19}& \ul{3.73}& \ul{2.47}& \ul{4.21}& \ul{2.22}& \ul{27.28}&\ul{19.31}\\
        \textbf{VAPO-7B (Ours)}& \bf4.60& \bf2.83& \bf2.97& \bf2.13& \bf3.78&\bf1.36& \bf26.48&\bf15.35\\
    \bottomrule
    \end{tabular}
    }
    \label{tab:main-SlideASR-Bench}
\end{table*}

\textbf{Implementation Details.} We fine-tune Qwen2.5-Omni (3B/7B) on the training set of SlideASR-S using VAPO for a total of 800 training steps. The training employs the AdamW~\cite{AadmW} optimizer (learning rate $1e^{-6}$, global batch size 32) on 4$\times$A100 GPUs. We set the group size to 4 and use a sampling temperature of 1.0 to encourage exploration during policy updates with a KL penalty coefficient of 0.01. The weights of the reward functions, $\lambda_1$ to $\lambda_4$, are all set to 1.

~\\
\textbf{Baselines \& Settings.} We benchmark mainstream LALMs: Qwen2-Audio~\cite{qwen2audio} and Mi-Dasheng~\cite{MiDashengLM} and OLLMs: MiniCPM-o-2.6~\cite{minicpmo} Qwen2.5-Omni~\cite{qwen25omni}, Qwen3-Omni~\cite{qwen3omni} on SlideSpeech~\cite{slidespeech} and SlideASR-Bench across three settings:

% \begin{enumerate}[leftmargin=*, noitemsep]
\begin{enumerate}[noitemsep]
    \item \textit{\textbf{Contextless}},  which only use audio as inputs;
    \item \textit{\textbf{Slide text as context}} (Pipeline), employing PaddleOCR~\cite{paddleocr} for textual context extraction following~\citet{chineselips};
    \item \textit{\textbf{Slide image as context}} (End-to-end).
\end{enumerate}

\noindent
We provide more evaluation details in Appendix~\ref{sec:app-prompts}. Additionally, we report results on the ChineseLips~\cite{chineselips} dataset in Appendix~\ref{sec:app-chineselips} to demonstrate the generalization of VAPO. 

~\\
\textbf{Metrics.} We utilize two sets of metrics consistent with prior works. For SlideSpeech~\cite{slidespeech}, we employ four metrics: WER, which measures overall transcription; U-WER, the unbiased WER on non-keyword segments for general transcription quality; B-WER, the biased WER focusing on keyword spans; and Recall, which is the percentage of correctly recognized keywords. Furthermore, for SlideASR-Bench, consistent with ContextASR-Bench~\cite{ContextASR-Bench}, we focus on three metrics: WER; NE-WER, the WER of the named entity portion; and NE-FNR, the False Negative Ratio of named entities, which measures the proportion of missed ground-truth entities. See Appendix~\ref{sec:app-metrics} for specific calculation details of the metrics.

\subsection{Main Results and Analysis}

\textbf{Results on SlideSpeech.} Table~\ref{tab:main-SlideSpeech} presents the results on the real-world SlideSpeech dataset. Ideally, visual context should enhance recognition; however, we observe a detrimental effect in baseline models. Most baselines exhibit performance degradation when incorporating slide information compared to the contextless setting. For instance, the Qwen3-Omni~\cite{qwen3omni} shows increased WER in both pipeline and end-to-end settings compared to using audio alone. In stark contrast, our VAPO method effectively leverages visual cues, achieving the best performance. VAPO-7B reaches a WER of 10.31 and a Recall of 97.32, significantly outperforming  the Qwen3-Omni baseline and the previous SOTA MaLa-ASR~\cite{MaLa-ASR}.

~\\
\textbf{Results on SlideASR-Bench.} Table~\ref{tab:main-SlideASR-Bench} reports the performance on our benchmark. Two critical observations highlight the superiority of VAPO: \\
\textbf{1) Audio-only models struggle with specialized entities.} On the challenging SlideASR-R subset, even the strongest audio-only model (Qwen3-Omni) suffers from a high NE-FNR of 41.09, underscoring the necessity of visual presentation context for domain-specific terms. \\
\textbf{2) Naive visual integration leads to catastrophic failure.} On SlideASR-S, baseline OLLMs fail to utilize visual context effectively. Instead of improving, they suffer from severe visual interference. For example, Qwen3-Omni in the end-to-end setting yields an exploded WER of 101.45 and a worsened NE-WER of 59.64 (vs. 14.61 in audio-only), indicating that the model is hallucinating slide content rather than transcribing speech. Conversely, on SlideASR-S (en), VAPO-3B achieves a remarkable WER of 4.90. Furthermore, on the SlideASR-R, VAPO-7B reduces the NE-FNR from the best baseline's 28.22 to 15.35. These results confirm that VAPO successfully enforces the ``\textit{Look-then-Listen}'' paradigm, accurately anchoring visual entities without succumbing to interference. We present case study in Appendix~\ref{sec:app-case-study} to provide visual comparison.

\begin{table}[!t]\small
    \caption{Ablation results of rewards on SlideASR-R.}

    \centering
    \begin{tabular}{ccc|cc}
    \toprule
              \fontsize{8.}{9}{\textbf{\makecell{ASR\\Reward}}}& \fontsize{8.2}{9}{\textbf{\makecell{OCR\\Reward}}}& \fontsize{8.2}{9}{\textbf{\makecell{VA\\Reward}}}& \fontsize{8.2}{9}{\textbf{NE-WER}} $\downarrow$& \fontsize{8.2}{9}{\textbf{NE-FNR}}$\downarrow$\\ \midrule
         
              \rowcolor{gray!13}
              \multicolumn{5}{c}{Qwen2.5-Omni-3B}\\
              \ding{56}& \ding{56}& \ding{56}& 49.00 &53.47 \\
              \ding{52}& \ding{56}& \ding{56}& 37.23 &31.19 \\
              \ding{52}& \ding{52}& \ding{56}& 29.97 &22.28 \\
              \ding{52}& \ding{52}& \ding{52}& \bf27.28&\bf19.31\\ \midrule
              
              \rowcolor{gray!13}
              \multicolumn{5}{c}{Qwen2.5-Omni-7B}\\
              \ding{56}& \ding{56}& \ding{56}& 41.77 &35.15 \\
              \ding{52}& \ding{56}& \ding{56}& 28.63 &20.30 \\
              \ding{52}& \ding{52}& \ding{56}& 26.75 &18.32 \\
              \ding{52}& \ding{52}& \ding{52}& \bf26.48&\bf15.35\\

 \bottomrule
    \end{tabular}
    \label{tab:ablation-function}
\end{table}

\begin{table}[!t]\small\setlength{\tabcolsep}{4.9pt}
    \centering
    \caption{Sensitivity analysis of reward weights ($\lambda_1$:$\lambda_2$:$\lambda_3$:$\lambda_4$) on SlideASR-S. The balanced 1:1:1:1 strategy achieves the best robustness across metrics (N-W: NE-WER, N-F: NE-FNR).}
    \begin{tabular}{l p{.7cm} p{.7cm} p{.7cm} p{.7cm} p{.7cm} p{.7cm}}
    \toprule
        \multirow{2}{*}{\textbf{Weights}} & \multicolumn{3}{c}{\textbf{SlideASR-S (en)}}& \multicolumn{3}{c}{\textbf{SlideASR-S (zh)}}\\
        \cmidrule(lr){2-4} \cmidrule(lr){5-7}
        & \fontsize{8}{9}{\textbf{WER}} & \fontsize{8}{9}{\textbf{N-W}}& \fontsize{8}{9}{\textbf{N-F}}& \fontsize{8}{9}{\textbf{WER}}& \fontsize{8}{9}{\textbf{N-W}}& \fontsize{8}{9}{\textbf{N-F}}\\ \midrule
         1:1:1:1& \bf4.90& \bf3.19& \bf3.73& \bf2.47&  \bf4.21& 2.22 \\
         1:1:1:2& 5.27 & 3.34 & 3.78 & 2.50 & 4.30 & 2.09 \\
         1:1:2:1& 5.32 & 4.12 & 3.91 & 2.48 & 4.38 & 2.09 \\
         1:2:1:1& 5.17 & 3.45 & 3.80 & 2.51 & 4.23 & \bf1.99\\
    \bottomrule
    \end{tabular}
    \label{tab:ablation-lmabda}
\end{table}

\subsection{Ablation Study}

\noindent\textbf{Ablation on Reward Functions.} Table~\ref{tab:ablation-function} details the stepwise contributions on SlideASR-R. Compared to the baseline, introducing the ASR Reward yields the most significant boost, drastically reducing NE-WER by stabilizing generation. Adding the OCR Reward further refines the visual prior quality. Finally, incorporating the Visual Anchoring Reward achieves the best performance. This confirms that while ASR and OCR guarantees individual modality perception, the VA reward is essential for effective referencing. We provide additional ablation results in Appendix~\ref{sec:app-more-ablation}.  

~\\
\textbf{Sensitivity to Reward Weights.} To investigate hyperparameter sensitivity, we evaluate varying weight configurations on SlideASR-S, as detailed in Table~\ref{tab:ablation-lmabda}. The results confirm that the balanced 1:1:1:1 scheme yields the most robust overall performance. We observe an inherent trade-off: doubling the VA reward ($\lambda_4$=2) enhances entity recall on the Chinese subset (improving NE-FNR to 2.09), but at the cost of increased overall WER due to over-aggressive anchoring. Conversely, double the ASR reward ($\lambda_3$=2) proves counterproductive, suppressing the model's reliance on visual cues and significantly degrading NE-WER (e.g., rising from 3.19 to 4.12 on the English subset). Consequently, we adopt the equal weighting strategy for its simplicity and effectiveness.

\subsection{Robustness against Mismatched Slides}
\begin{figure}[!t]
\setlength{\abovecaptionskip}{1pt}
    \centering
    \includegraphics[width=1\linewidth]{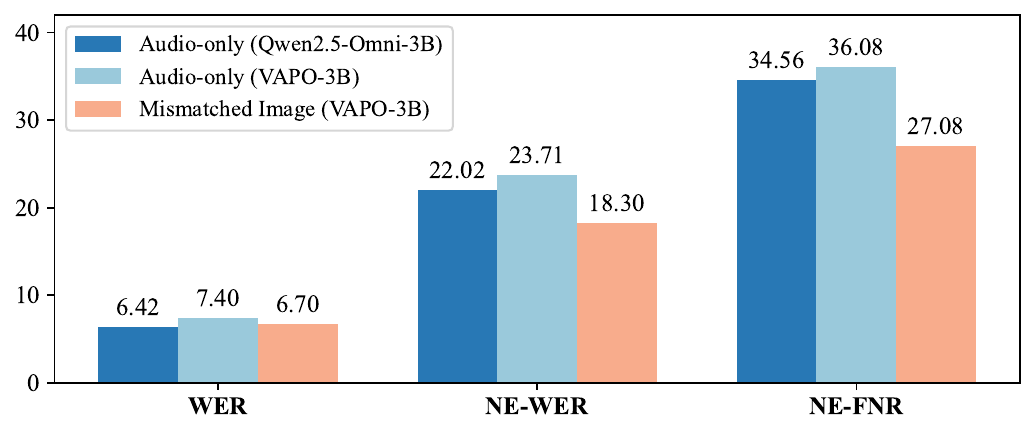}
    \caption{Robustness analysis of VAPO-3B under mismatched visual context on SlideASR-S.}
    \label{fig:mismatch}
\end{figure}

To verify that VAPO performs valid visual anchoring rather than blindly copying slide content, we evaluated the model's performance when fed with randomly mismatched slide images. As illustrated in Fig.~\ref{fig:mismatch}, the model exhibits strong robustness: the overall WER remains stable at 6.70\%, which is comparable to the audio-only baselines, including Qwen2.5-Omni-3B (6.42\%) and the audio-only variant of VAPO-3B (7.40\%). This indicates that VAPO effectively mitigates visual interference, safely falling back to auditory perception when visual context is unreliable.

% \textbf{Robustness against Mismatched Visual Context.} To verify that VAPO performs valid visual anchoring rather than blindly cpoying slide content, we evaluated the model's performance when fed with randomly mismatched slide images. As illustrated in Fig.~\ref{fig:mismatch}, the model exhibits strong robustness: the overall WER remains stable at 6.70\%, comparable to the Audio-only baseline of 6.42\%. This indicates that VAPO effectively mitigates visual interference, safely falling back to auditory perception when visual context are unreliable.

\begin{table}[!t]\small
    \centering
    \caption{Comparison between VAPO and SFT baselines on SlideASR-S. \textbf{"SFT w/o think"} is trained to directly predict the transcription, whereas \textbf{"SFT w/ think"} is supervised using the structured inference chain.}
    \begin{tabular}{lccc} \toprule
         \textbf{Model}& \fontsize{8.5}{9}{\textbf{WER}}$\downarrow$& \fontsize{8.5}{9}{\textbf{NE-WER}} $\downarrow$& \fontsize{8.5}{9}{\textbf{NE-FNR}}$\downarrow$\\ \midrule
         \rowcolor{gray!13}Qwen2.5-Omni-3B& & &\\ 
         \textbf{+ SFT w/o think}& 8.44 & 8.18 & 9.30 \\
 \textbf{+ SFT w/ think}& 6.60 & 6.47 &7.10 \\
         \textbf{+ VAPO}& \bf3.88& \bf3.84& \bf3.00\\ \midrule
         \rowcolor{gray!13}Qwen2.5-Omni-7B & & &\\ 
         \textbf{+ SFT w/o think}& 10.58& 10.51& 12.77\\
 \textbf{+ SFT w/ think}& 6.73 & 5.56 &6.53 \\
         \textbf{+ VAPO}& \bf3.37& \bf3.07& \bf2.00\\
    \bottomrule
    \end{tabular}
    \label{tab:ablation-sft}
\end{table}

\begin{figure*}
    \includegraphics[width=.95\linewidth]{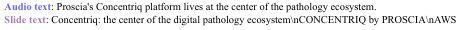}
    \includegraphics[width=.215\linewidth]{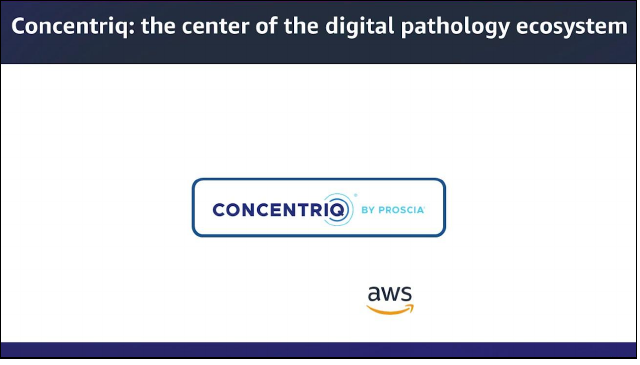}
    \hspace{3pt}
    \includegraphics[width=.765\linewidth]{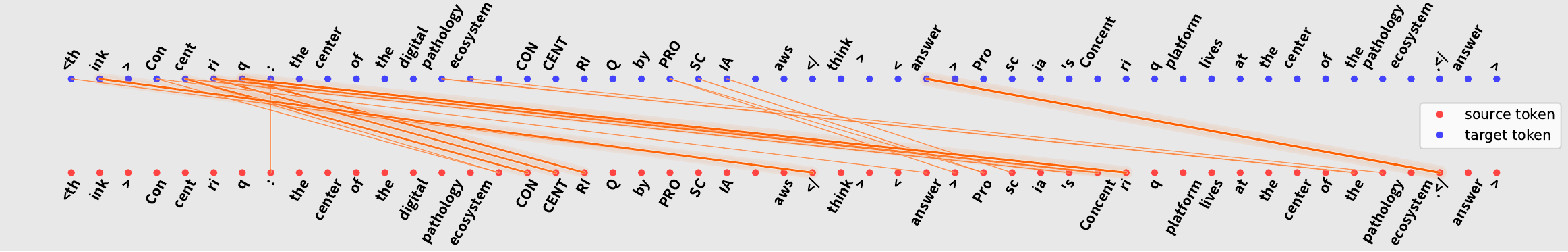}
    
    \caption{Attention visualization of VAPO-7B. The dense orange connections demonstrate the visual anchoring mechanism, showing that the model explicitly attends to extracted entities in \textit{<think>} to guide the final transcription.}
    \label{fig:attention}
\end{figure*}

\subsection{Impact of Training Paradigm: VAPO vs. Supervised Fine-Tuning (SFT)}

To disentangle the contributions of the structured format from the reinforcement learning (RL) optimization, we compare three training settings on SlideASR-S: SFT w/o think (direct mapping), SFT w/ think (structured but SFT-only), and VAPO. Results in Table~\ref{tab:ablation-sft} highlights two key findings: \\
\textbf{1) Efficacy of the Inference Chain.} Introducing a structured inference chain yields significant gains, reducing the WER of Qwen2.5-Omni-7B from 10.58 to 6.73. This confirms that temporally decoupling the ``Look'' and ``Listen'' phases inherently mitigates visual interference. \\
\textbf{2) Necessity of RL.} VAPO further lowers the WER to 3.37. This indicates that while SFT teaches the format, it fails to establish a sufficient connection between \textit{<think>}  and \textit{<answer>}. VAPO, driven by the multi-dimensional rewards, ensures the \textit{<think>} block is effectively utilized for the final transcription.

\subsection{Attention Visualization}

To empirically validate the effectiveness of the visual anchoring mechanism, we visualize the attention weights of VAPO-7B on a sample from SlideSpeech, as shown in Fig.~\ref{fig:attention}. The visualization reveals a distinct semantic reference pattern: during the generation of the \textit{<answer>} block, the model heavily attends to the corresponding extracted text within the \textit{<think>} block. 

Specifically, take the proper noun ``\textit{Concentriq}'' as an example. After generating the token ``\textit{Concent}'' in \textit{<answer>}, the model pays significant attention to the ``\textit{ri}'' token in \textit{<think>} and subsequently generates it. It then refers to the ``\textit{q}'' token in the \textit{<think>} block, enabling accurate and complete transcription of ``\textit{Concentriq}''. A similar process occurs for the entity ``\textit{proscia}''. This qualitative evidence confirms that VAPO successfully enforces the ``\textit{Look-then-Listen}'' paradigm, where the model explicitly references the visual context prior. We provide more cases of attention visualization in Appendix~\ref{sec:app-more-attention-visualization}.

\section{Related Works}
\textbf{Contextual ASR.} The objective of contextual ASR is to incorporate contextual information, including domain labels, entity lists, and conversational history, into the speech recognition system in order to improve the recognition accuracy of  domain-specific terminology~\cite{SeedASR, contextasr_with_Retrieval, ContextualASR_via_Fusion}.  Besides textual information, researchers have focused on leveraging visual  information to enhance the performance of  ASR models. For example, integrating lip movement information during the recognition process~\cite{AutoAVSR, WhisperFlamingo, AVHubert}. This study focuses on the SlideASR task~\cite{chineselips, slidespeech, slideavsr}, which involves utilizing slide content as contextual information to support the model, given that slides in presentation scenarios generally contain information closely related to the spoken content. Most existing methods  for SlideASR are based on the pipeline paradigm~\cite{slideavsr, chineselips, slidespeech, LCB-Net, MaLa-ASR}, which results in relatively complex systems. The objective of this study is to accomplish the task using an end-to-end approach. 

~\\
\textbf{Omni-modal Large Language Models.}  Recently, OLLMs~\cite{hurst2024gpt, VITA1.5, minicpmo, qwen25omni, Megrez, selfkd, baichuan-omni-1.5} have emerged, integrating vision, audio, and text by aligning their encoders during training for end-to-end processing. Models such as MiniCPM-o \cite{minicpmo} and Qwen2.5-Omni~\cite{qwen25omni} have demonstrated strong multimodal performance. Benefiting from the unified modeling capability across visual and audio modalities, they are expected to solve the SlideASR task in an end-to-end manner. However, in practical scenarios, the models exhibit visual interference, for instance, they sometimes reproduce the textual content from the slides instead of generating the expected speech transcription. 

~\\
\textbf{Chain-of-Thought Reasoning.} CoT reasoning is a breakthrough approach that enhances the reasoning capabilities of LLMs. Recent works~\cite{o1, deepseekr1} have shown that by using reinforcement learning algorithms~\cite{deepseekmath, dpo, ppo} to encourage models to generate intermediate reasoning steps before producing the final answer, performance on tasks involving arithmetic, commonsense, and symbolic reasoning can be significantly enhanced.
 This paradigm has also been extended to the multimodal domain~\cite{visualcot, llavaCot, audioCoT, MindWithEyes, SoundMind}, demonstrating the general effectiveness of making reasoning processes explicit. 

\section{Conclusion}
In this work, we uncover a critical \textit{Visual Interference} phenomenon in OLLMs applied to the SlideASR task, where models succumb to modality dominance and ignore auditory inputs. To address this, we introduce Visually-Anchored Policy Optimization (VAPO), a novel post-training method that enforces a human-like ``\textit{Look-then-Listen}'' inference chain. By leveraging a structured \textit{<think><answer>} format and multi-dimensional reward optimization, VAPO effectively decouples visual perception from auditory transcription. Furthermore, we contribute SlideASR-Bench, a comprehensive benchmark designed to address the data scarcity in entity-rich scenarios. Extensive experiments demonstrate that VAPO not only eliminates visual interference but also sets a new state-of-the-art performance, particularly in recognizing specialized domain entities.

\section{Limitations}
While VAPO demonstrates significant improvements, this work has several limitations.

\noindent
\textbf{Task Generalization.} Our current approach is highly specialized for leveraging textual information from presentation slides and does not incorporate other visual cues, such as images of entities (e.g., pictures of specific drugs). In the future, we will adapt our ``Look-then-Listen'' paradigm to handle more diverse multimodal environments, where various visual elements play a crucial role. 

~\\
\textbf{Real-World Robustness.} While our training relies on the synthetic SlideASR-S dataset, we have validated its effectiveness on three real-world datasets, SlideSpeech, ChineseLips and SlideASR-R. Nonetheless, a subtle domain gap may still exist, as synthetic slides may not fully capture the stylistic diversity and visual noise (e.g., complex diagrams, low-quality images) of all real-world presentations.

~\\
\textbf{Inference Efficiency.} The structured reasoning process of VAPO introduces a computational overhead, resulting in higher inference latency compared to models of the same size (detailed in Appendix~\ref{sec:app-latency}). This makes VAPO most suitable for offline applications where accuracy is critical. However, this trade-off between latency and accuracy is often acceptable for offline transcription tasks where precision is paramount. We will explore strategies such as model distillation in our future research to improve efficiency, enabling its use in real-time applications.

\section{Ethical Considerations}

The primary societal benefit of our work is enhancing accessibility by improving the transcription accuracy of specialized terms, which can significantly aid individuals who are deaf or hard of hearing. However, this technology must be deployed with caution in high-stakes settings, such as medical transcription, where errors could lead to serious consequences. We advocate for responsible development and believe that human oversight is essential for any critical applications.

\section{Acknowledgments}

This work is supported by the National Key R\&D Program of China (No. 2023YFC3310700), the Open Funding Programs of State Key Laboratory of AI Safety, the National Natural Science Foundation of China (No. 2576030), and the Beijing Nova Program under Grant 20250484899.

\bibliography{acl_latex}

@inproceedings{radford2023whisper,
  author       = {Alec Radford and
                  Jong Wook Kim and
                  Tao Xu and
                  Greg Brockman and
                  Christine McLeavey and
                  Ilya Sutskever},
  editor       = {Andreas Krause and
                  Emma Brunskill and
                  Kyunghyun Cho and
                  Barbara Engelhardt and
                  Sivan Sabato and
                  Jonathan Scarlett},
  title        = {Robust Speech Recognition via Large-Scale Weak Supervision},
  booktitle    = {International Conference on Machine Learning, {ICML} 2023, 23-29 July
                  2023, Honolulu, Hawaii, {USA}},
  series       = {Proceedings of Machine Learning Research},
  volume       = {202},
  pages        = {28492--28518},
  publisher    = {{PMLR}},
  year         = {2023},
  url          = {https://proceedings.mlr.press/v202/radford23a.html},
  bibsource    = {dblp computer science bibliography, https://dblp.org}
}

@inproceedings{sinhamahapatra2025slides,
  title={Do Slides Help? Multi-modal Context for Automatic Transcription of Conference Talks},
  author={Sinhamahapatra, Supriti and Niehues, Jan},
  booktitle={Proceedings of the 2025 Conference on Empirical Methods in Natural Language Processing},
  pages={16111--16121},
  year={2025}
}

@article{deep-avsr,
  author       = {Triantafyllos Afouras and
                  Joon Son Chung and
                  Andrew W. Senior and
                  Oriol Vinyals and
                  Andrew Zisserman},
  title        = {Deep Audio-Visual Speech Recognition},
  journal      = {{IEEE} Trans. Pattern Anal. Mach. Intell.},
  volume       = {44},
  number       = {12},
  pages        = {8717--8727},
  year         = {2022},
  url          = {https://doi.org/10.1109/TPAMI.2018.2889052},
  doi          = {10.1109/TPAMI.2018.2889052},
  bibsource    = {dblp computer science bibliography, https://dblp.org}
}

@inproceedings{conformers-avsr,
  author       = {Pingchuan Ma and
                  Stavros Petridis and
                  Maja Pantic},
  title        = {End-To-End Audio-Visual Speech Recognition with Conformers},
  booktitle    = {{IEEE} International Conference on Acoustics, Speech and Signal Processing,
                  {ICASSP} 2021, Toronto, ON, Canada, June 6-11, 2021},
  pages        = {7613--7617},
  publisher    = {{IEEE}},
  year         = {2021},
  url          = {https://doi.org/10.1109/ICASSP39728.2021.9414567},
  doi          = {10.1109/ICASSP39728.2021.9414567},
  bibsource    = {dblp computer science bibliography, https://dblp.org}
}

@inproceedings{slidespeech,
  author       = {Haoxu Wang and
                  Fan Yu and
                  Xian Shi and
                  Yuezhang Wang and
                  Shiliang Zhang and
                  Ming Li},
  title        = {SlideSpeech: {A} Large Scale Slide-Enriched Audio-Visual Corpus},
  booktitle    = {{IEEE} International Conference on Acoustics, Speech and Signal Processing,
                  {ICASSP} 2024, Seoul, Republic of Korea, April 14-19, 2024},
  pages        = {11076--11080},
  publisher    = {{IEEE}},
  year         = {2024},
  url          = {https://doi.org/10.1109/ICASSP48485.2024.10448079},
  doi          = {10.1109/ICASSP48485.2024.10448079},
  bibsource    = {dblp computer science bibliography, https://dblp.org}
}

@article{slideavsr,
  author       = {Hao Wang and
                  Shuhei Kurita and
                  Shuichiro Shimizu and
                  Daisuke Kawahara},
  title        = {SlideAVSR: {A} Dataset of Paper Explanation Videos for Audio-Visual
                  Speech Recognition},
  journal      = {Proceedings of the 3rd Workshop on Advances in Language and Vision Research (ALVR)},
  year         = {2024},
  url          = {https://aclanthology.org/2024.alvr-1.11/},
  pages        = {129--137}
}

@article{chineselips,
  author       = {Jinghua Zhao and
                  Yuhang Jia and
                  Shiyao Wang and
                  Jiaming Zhou and
                  Hui Wang and
                  Yong Qin},
  title        = {Chinese-LiPS: {A} Chinese audio-visual speech recognition dataset
                  with Lip-reading and Presentation Slides},
  journal      = {CoRR},
  volume       = {abs/2504.15066},
  year         = {2025},
  url          = {https://doi.org/10.48550/arXiv.2504.15066},
  doi          = {10.48550/ARXIV.2504.15066},
  eprinttype    = {arXiv},
  eprint       = {2504.15066},
  bibsource    = {dblp computer science bibliography, https://dblp.org}
}

@article{qwen2audio,
  author       = {Yunfei Chu and
                  Jin Xu and
                  Qian Yang and
                  Haojie Wei and
                  Xipin Wei and
                  Zhifang Guo and
                  Yichong Leng and
                  Yuanjun Lv and
                  Jinzheng He and
                  Junyang Lin and
                  Chang Zhou and
                  Jingren Zhou},
  title        = {Qwen2-Audio Technical Report},
  journal      = {CoRR},
  volume       = {abs/2407.10759},
  year         = {2024},
  url          = {https://doi.org/10.48550/arXiv.2407.10759},
  doi          = {10.48550/ARXIV.2407.10759},
  eprinttype    = {arXiv},
  eprint       = {2407.10759},
  bibsource    = {dblp computer science bibliography, https://dblp.org}
}

@article{MiDashengLM,
  author       = {Heinrich Dinkel and
                  Gang Li and
                  Jizhong Liu and
                  Jian Luan and
                  Yadong Niu and
                  Xingwei Sun and
                  Tianzi Wang and
                  Qiyang Xiao and
                  Junbo Zhang and
                  Jiahao Zhou},
  title        = {MiDashengLM: Efficient Audio Understanding with General Audio Captions},
  journal      = {CoRR},
  volume       = {abs/2508.03983},
  year         = {2025},
  url          = {https://doi.org/10.48550/arXiv.2508.03983},
  doi          = {10.48550/ARXIV.2508.03983},
  eprinttype    = {arXiv},
  eprint       = {2508.03983},
  bibsource    = {dblp computer science bibliography, https://dblp.org}
}

@article{qwen25omni,
  author       = {Jin Xu and
                  Zhifang Guo and
                  Jinzheng He and
                  Hangrui Hu and
                  Ting He and
                  Shuai Bai and
                  Keqin Chen and
                  Jialin Wang and
                  Yang Fan and
                  Kai Dang and
                  Bin Zhang and
                  Xiong Wang and
                  Yunfei Chu and
                  Junyang Lin},
  title        = {Qwen2.5-Omni Technical Report},
  journal      = {CoRR},
  volume       = {abs/2503.20215},
  year         = {2025},
  url          = {https://doi.org/10.48550/arXiv.2503.20215},
  doi          = {10.48550/ARXIV.2503.20215},
  eprinttype    = {arXiv},
  eprint       = {2503.20215},
  bibsource    = {dblp computer science bibliography, https://dblp.org}
}

@article{qwen3omni,
  title={Qwen3-Omni Technical Report},
  author={Xu, Jin and Guo, Zhifang and Hu, Hangrui and Chu, Yunfei and Wang, Xiong and He, Jinzheng and Wang, Yuxuan and Shi, Xian and He, Ting and Zhu, Xinfa and others},
  journal      = {CoRR},
  volume       = {abs/2509.17765},
  year         = {2025},
  url          = {https://arxiv.org/abs/2509.17765},
}

@article{minicpmo,
  author       = {Yuan Yao and
                  Tianyu Yu and
                  Ao Zhang and
                  Chongyi Wang and
                  Junbo Cui and
                  Hongji Zhu and
                  Tianchi Cai and
                  Haoyu Li and
                  Weilin Zhao and
                  Zhihui He and
                  Qianyu Chen and
                  Huarong Zhou and
                  Zhensheng Zou and
                  Haoye Zhang and
                  Shengding Hu and
                  Zhi Zheng and
                  Jie Zhou and
                  Jie Cai and
                  Xu Han and
                  Guoyang Zeng and
                  Dahai Li and
                  Zhiyuan Liu and
                  Maosong Sun},
  title        = {MiniCPM-V: {A} {GPT-4V} Level {MLLM} on Your Phone},
  journal      = {CoRR},
  volume       = {abs/2408.01800},
  year         = {2024},
  url          = {https://doi.org/10.48550/arXiv.2408.01800},
  doi          = {10.48550/ARXIV.2408.01800},
  eprinttype    = {arXiv},
  eprint       = {2408.01800},
  bibsource    = {dblp computer science bibliography, https://dblp.org}
}

@article{Megrez,
  author       = {Boxun Li and
                  Yadong Li and
                  Zhiyuan Li and
                  Congyi Liu and
                  Weilin Liu and
                  Guowei Niu and
                  Zheyue Tan and
                  Haiyang Xu and
                  Zhuyu Yao and
                  Tao Yuan and
                  Dong Zhou and
                  Yueqing Zhuang and
                  Shengen Yan and
                  Guohao Dai and
                  Yu Wang},
  title        = {Megrez-Omni Technical Report},
  journal      = {CoRR},
  volume       = {abs/2502.15803},
  year         = {2025},
  url          = {https://doi.org/10.48550/arXiv.2502.15803},
  doi          = {10.48550/ARXIV.2502.15803},
  eprinttype    = {arXiv},
  eprint       = {2502.15803},
  bibsource    = {dblp computer science bibliography, https://dblp.org}
}

@article{SeedASR,
  author       = {Ye Bai and
                  Jingping Chen and
                  Jitong Chen and
                  Wei Chen and
                  Zhuo Chen and
                  Chuang Ding and
                  Linhao Dong and
                  Qianqian Dong and
                  Yujiao Du and
                  Kepan Gao and
                  Lu Gao and
                  Yi Guo and
                  Minglun Han and
                  others},
  title        = {Seed-ASR: Understanding Diverse Speech and Contexts with LLM-based
                  Speech Recognition},
  journal      = {CoRR},
  volume       = {abs/2407.04675},
  year         = {2024},
  url          = {https://doi.org/10.48550/arXiv.2407.04675},
  doi          = {10.48550/ARXIV.2407.04675},
  eprinttype    = {arXiv},
  eprint       = {2407.04675},
  bibsource    = {dblp computer science bibliography, https://dblp.org}
}

@inproceedings{contextasr_with_Retrieval,
  author       = {Cihan Xiao and
                  Zejiang Hou and
                  Daniel Garcia{-}Romero and
                  Kyu J. Han},
  title        = {Contextual {ASR} with Retrieval Augmented Large Language Model},
  booktitle    = {2025 {IEEE} International Conference on Acoustics, Speech and Signal
                  Processing, {ICASSP} 2025, Hyderabad, India, April 6-11, 2025},
  pages        = {1--5},
  publisher    = {{IEEE}},
  year         = {2025},
  url          = {https://doi.org/10.1109/ICASSP49660.2025.10890057},
  doi          = {10.1109/ICASSP49660.2025.10890057},
  bibsource    = {dblp computer science bibliography, https://dblp.org}
}

@article{ContextualASR_via_Fusion,
  author       = {Shilin Zhou and
                  Zhenghua Li},
  title        = {Improving Contextual {ASR} via Multi-grained Fusion with Large Language
                  Models},
  journal      = {CoRR},
  volume       = {abs/2507.12252},
  year         = {2025},
  url          = {https://doi.org/10.48550/arXiv.2507.12252},
  doi          = {10.48550/ARXIV.2507.12252},
  eprinttype    = {arXiv},
  eprint       = {2507.12252},
  bibsource    = {dblp computer science bibliography, https://dblp.org}
}

@inproceedings{AutoAVSR,
  author       = {Pingchuan Ma and
                  Alexandros Haliassos and
                  Adriana Fernandez{-}Lopez and
                  Honglie Chen and
                  Stavros Petridis and
                  Maja Pantic},
  title        = {Auto-AVSR: Audio-Visual Speech Recognition with Automatic Labels},
  booktitle    = {{IEEE} International Conference on Acoustics, Speech and Signal Processing
                  {ICASSP} 2023, Rhodes Island, Greece, June 4-10, 2023},
  pages        = {1--5},
  publisher    = {{IEEE}},
  year         = {2023},
  url          = {https://doi.org/10.1109/ICASSP49357.2023.10096889},
  doi          = {10.1109/ICASSP49357.2023.10096889},
  bibsource    = {dblp computer science bibliography, https://dblp.org}
}

@inproceedings{WhisperFlamingo,
  author       = {Andrew Rouditchenko and
                  Yuan Gong and
                  Samuel Thomas and
                  Leonid Karlinsky and
                  Hilde Kuehne and
                  Rog{\'{e}}rio Feris and
                  James Glass},
  editor       = {Itshak Lapidot and
                  Sharon Gannot},
  title        = {Whisper-Flamingo: Integrating Visual Features into Whisper for Audio-Visual
                  Speech Recognition and Translation},
  booktitle    = {25th Annual Conference of the International Speech Communication Association,
                  Interspeech 2024, Kos, Greece, September 1-5, 2024},
  publisher    = {{ISCA}},
  year         = {2024},
  url          = {https://doi.org/10.21437/Interspeech.2024-322},
  doi          = {10.21437/INTERSPEECH.2024-322},
  bibsource    = {dblp computer science bibliography, https://dblp.org}
}

@inproceedings{AVHubert,
  author       = {Bowen Shi and
                  Wei{-}Ning Hsu and
                  Kushal Lakhotia and
                  Abdelrahman Mohamed},
  title        = {Learning Audio-Visual Speech Representation by Masked Multimodal Cluster
                  Prediction},
  booktitle    = {The Tenth International Conference on Learning Representations, {ICLR}
                  2022, Virtual Event, April 25-29, 2022},
  publisher    = {OpenReview.net},
  year         = {2022},
  url          = {https://openreview.net/forum?id=Z1Qlm11uOM},
  bibsource    = {dblp computer science bibliography, https://dblp.org}
}

@inproceedings{LCB-Net,
  author       = {Fan Yu and
                  Haoxu Wang and
                  Xian Shi and
                  Shiliang Zhang},
  title        = {LCB-Net: Long-Context Biasing for Audio-Visual Speech Recognition},
  booktitle    = {{IEEE} International Conference on Acoustics, Speech and Signal Processing,
                  {ICASSP} 2024, Seoul, Republic of Korea, April 14-19, 2024},
  pages        = {10621--10625},
  publisher    = {{IEEE}},
  year         = {2024},
  url          = {https://doi.org/10.1109/ICASSP48485.2024.10448106},
  doi          = {10.1109/ICASSP48485.2024.10448106},
  bibsource    = {dblp computer science bibliography, https://dblp.org}
}

@inproceedings{MaLa-ASR,
  author       = {Guanrou Yang and
                  Ziyang Ma and
                  Fan Yu and
                  Zhifu Gao and
                  Shiliang Zhang and
                  Xie Chen},
  editor       = {Itshak Lapidot and
                  Sharon Gannot},
  title        = {MaLa-ASR: Multimedia-Assisted LLM-Based {ASR}},
  booktitle    = {25th Annual Conference of the International Speech Communication Association,
                  Interspeech 2024, Kos, Greece, September 1-5, 2024},
  publisher    = {{ISCA}},
  year         = {2024},
  url          = {https://doi.org/10.21437/Interspeech.2024-488},
  doi          = {10.21437/INTERSPEECH.2024-488},
  bibsource    = {dblp computer science bibliography, https://dblp.org}
}

@article{VITA1.5,
  author       = {Chaoyou Fu and
                  Haojia Lin and
                  Xiong Wang and
                  Yifan Zhang and
                  Yunhang Shen and
                  Xiaoyu Liu and
                  Haoyu Cao and
                  Zuwei Long and
                  Heting Gao and
                  Ke Li and
                  Long Ma and
                  Xiawu Zheng and
                  Rongrong Ji and
                  Xing Sun and
                  Caifeng Shan and
                  Ran He},
  title        = {{VITA-1.5:} Towards GPT-4o Level Real-Time Vision and Speech Interaction},
  journal      = {CoRR},
  volume       = {abs/2501.01957},
  year         = {2025},
  url          = {https://doi.org/10.48550/arXiv.2501.01957},
  doi          = {10.48550/ARXIV.2501.01957},
  eprinttype    = {arXiv},
  eprint       = {2501.01957},
  bibsource    = {dblp computer science bibliography, https://dblp.org}
}

@inproceedings{selfkd,
  author       = {Rui Hu and
                  Delai Qiu and
                  Shuyu Wei and
                  Jiaming Zhang and
                  Yining Wang and
                  Shengping Liu and
                  Jitao Sang},
  editor       = {Wanxiang Che and
                  Joyce Nabende and
                  Ekaterina Shutova and
                  Mohammad Taher Pilehvar},
  title        = {Investigating and Enhancing Vision-Audio Capability in Omnimodal Large
                  Language Models},
  booktitle    = {Findings of the Association for Computational Linguistics, {ACL} 2025,
                  Vienna, Austria, July 27 - August 1, 2025},
  pages        = {7452--7463},
  publisher    = {Association for Computational Linguistics},
  year         = {2025},
  url          = {https://aclanthology.org/2025.findings-acl.389/},
  bibsource    = {dblp computer science bibliography, https://dblp.org}
}

@article{baichuan-omni-1.5,
  author={Li, Yadong and Liu, Jun and Zhang, Tao and Chen, Song and Li, Tianpeng and Li, Zehuan and Liu, Lijun and Ming, Lingfeng and Dong, Guosheng and Pan, Da and others},
  title        = {Baichuan-Omni-1.5 Technical Report},
  journal      = {CoRR},
  volume       = {abs/2501.15368},
  year         = {2025},
  url          = {https://doi.org/10.48550/arXiv.2501.15368},
  doi          = {10.48550/ARXIV.2501.15368},
  eprinttype    = {arXiv},
  eprint       = {2501.15368},
  bibsource    = {dblp computer science bibliography, https://dblp.org}
}

@article{o1,
author={Jaech, Aaron and Kalai, Adam and Lerer, Adam and Richardson, Adam and El-Kishky, Ahmed and Low, Aiden and Helyar, Alec and Madry, Aleksander and Beutel, Alex and Carney, Alex and others},
  title        = {OpenAI o1 System Card},
  journal      = {CoRR},
  volume       = {abs/2412.16720},
  year         = {2024},
  url          = {https://doi.org/10.48550/arXiv.2412.16720},
  doi          = {10.48550/ARXIV.2412.16720},
  eprinttype    = {arXiv},
  eprint       = {2412.16720},
  bibsource    = {dblp computer science bibliography, https://dblp.org}
}

@article{deepseekr1,
  author={Guo, Daya and Yang, Dejian and Zhang, Haowei and Song, Junxiao and Zhang, Ruoyu and Xu, Runxin and Zhu, Qihao and Ma, Shirong and Wang, Peiyi and Bi, Xiao and others},
  title        = {DeepSeek-R1: Incentivizing Reasoning Capability in LLMs via Reinforcement
                  Learning},
  journal      = {CoRR},
  volume       = {abs/2501.12948},
  year         = {2025},
  url          = {https://doi.org/10.48550/arXiv.2501.12948},
  doi          = {10.48550/ARXIV.2501.12948},
  eprinttype    = {arXiv},
  eprint       = {2501.12948},
  bibsource    = {dblp computer science bibliography, https://dblp.org}
}

@article{deepseekmath,
  author       = {Zhihong Shao and
                  Peiyi Wang and
                  Qihao Zhu and
                  Runxin Xu and
                  Junxiao Song and
                  Mingchuan Zhang and
                  Y. K. Li and
                  Y. Wu and
                  Daya Guo},
  title        = {DeepSeekMath: Pushing the Limits of Mathematical Reasoning in Open
                  Language Models},
  journal      = {CoRR},
  volume       = {abs/2402.03300},
  year         = {2024},
  url          = {https://doi.org/10.48550/arXiv.2402.03300},
  doi          = {10.48550/ARXIV.2402.03300},
  eprinttype    = {arXiv},
  eprint       = {2402.03300},
  bibsource    = {dblp computer science bibliography, https://dblp.org}
}

@inproceedings{dpo,
  author       = {Rafael Rafailov and
                  Archit Sharma and
                  Eric Mitchell and
                  Christopher D. Manning and
                  Stefano Ermon and
                  Chelsea Finn},
  editor       = {Alice Oh and
                  Tristan Naumann and
                  Amir Globerson and
                  Kate Saenko and
                  Moritz Hardt and
                  Sergey Levine},
  title        = {Direct Preference Optimization: Your Language Model is Secretly a
                  Reward Model},
  booktitle    = {Advances in Neural Information Processing Systems 36: Annual Conference
                  on Neural Information Processing Systems 2023, NeurIPS 2023, New Orleans,
                  LA, USA, December 10 - 16, 2023},
  year         = {2023},
  url          = {http://papers.nips.cc/paper\_files/paper/2023/hash/a85b405ed65c6477a4fe8302b5e06ce7-Abstract-Conference.html},
  bibsource    = {dblp computer science bibliography, https://dblp.org}
}

@article{ppo,
  author       = {John Schulman and
                  Filip Wolski and
                  Prafulla Dhariwal and
                  Alec Radford and
                  Oleg Klimov},
  title        = {Proximal Policy Optimization Algorithms},
  journal      = {CoRR},
  volume       = {abs/1707.06347},
  year         = {2017},
  url          = {http://arxiv.org/abs/1707.06347},
  eprinttype    = {arXiv},
  eprint       = {1707.06347},
  bibsource    = {dblp computer science bibliography, https://dblp.org}
}

@inproceedings{visualcot,
  author       = {Hao Shao and
                  Shengju Qian and
                  Han Xiao and
                  Guanglu Song and
                  Zhuofan Zong and
                  Letian Wang and
                  Yu Liu and
                  Hongsheng Li},
  editor       = {Amir Globersons and
                  Lester Mackey and
                  Danielle Belgrave and
                  Angela Fan and
                  Ulrich Paquet and
                  Jakub M. Tomczak and
                  Cheng Zhang},
  title        = {Visual CoT: Advancing Multi-Modal Language Models with a Comprehensive
                  Dataset and Benchmark for Chain-of-Thought Reasoning},
  booktitle    = {Advances in Neural Information Processing Systems 38: Annual Conference
                  on Neural Information Processing Systems 2024, NeurIPS 2024, Vancouver,
                  BC, Canada, December 10 - 15, 2024},
  year         = {2024},
  url          = {http://papers.nips.cc/paper\_files/paper/2024/hash/0ff38d72a2e0aa6dbe42de83a17b2223-Abstract-Datasets\_and\_Benchmarks\_Track.html},
  bibsource    = {dblp computer science bibliography, https://dblp.org}
}

@article{llavaCot,
  author       = {Guowei Xu and
                  Peng Jin and
                  Hao Li and
                  Yibing Song and
                  Lichao Sun and
                  Li Yuan},
  title        = {LLaVA-CoT: Let Vision Language Models Reason Step-by-Step},
  journal      = {CoRR},
  volume       = {abs/2411.10440},
  year         = {2024},
  url          = {https://doi.org/10.48550/arXiv.2411.10440},
  doi          = {10.48550/ARXIV.2411.10440},
  eprinttype    = {arXiv},
  eprint       = {2411.10440},
  bibsource    = {dblp computer science bibliography, https://dblp.org}
}

@article{audioCoT,
  author       = {Ziyang Ma and
                  Zhuo Chen and
                  Yuping Wang and
                  Eng Siong Chng and
                  Xie Chen},
  title        = {Audio-CoT: Exploring Chain-of-Thought Reasoning in Large Audio Language
                  Model},
  journal      = {CoRR},
  volume       = {abs/2501.07246},
  year         = {2025},
  url          = {https://doi.org/10.48550/arXiv.2501.07246},
  doi          = {10.48550/ARXIV.2501.07246},
  eprinttype    = {arXiv},
  eprint       = {2501.07246},
  bibsource    = {dblp computer science bibliography, https://dblp.org}
}

@article{SoundMind,
  author       = {Xingjian Diao and
                  Chunhui Zhang and
                  Keyi Kong and
                  Weiyi Wu and
                  Chiyu Ma and
                  Zhongyu Ouyang and
                  Peijun Qing and
                  Soroush Vosoughi and
                  Jiang Gui},
  title        = {SoundMind: RL-Incentivized Logic Reasoning for Audio-Language Models},
  journal      = {CoRR},
  volume       = {abs/2506.12935},
  year         = {2025},
  url          = {https://doi.org/10.48550/arXiv.2506.12935},
  doi          = {10.48550/ARXIV.2506.12935},
  eprinttype    = {arXiv},
  eprint       = {2506.12935},
  bibsource    = {dblp computer science bibliography, https://dblp.org}
}

@article{ContextASR-Bench,
  author       = {He Wang and
                  Linhan Ma and
                  Dake Guo and
                  Xiong Wang and
                  Lei Xie and
                  Jin Xu and
                  Junyang Lin},
  title        = {ContextASR-Bench: {A} Massive Contextual Speech Recognition Benchmark},
  journal      = {CoRR},
  volume       = {abs/2507.05727},
  year         = {2025},
  url          = {https://doi.org/10.48550/arXiv.2507.05727},
  doi          = {10.48550/ARXIV.2507.05727},
  eprinttype    = {arXiv},
  eprint       = {2507.05727},
  bibsource    = {dblp computer science bibliography, https://dblp.org}
}

@article{CosyVoice2,
  author       = {Zhihao Du and
                  Yuxuan Wang and
                  Qian Chen and
                  Xian Shi and
                  Xiang Lv and
                  Tianyu Zhao and
                  Zhifu Gao and
                  Yexin Yang and
                  Changfeng Gao and
                  Hui Wang and
                  Fan Yu and
                  Huadai Liu and
                  Zhengyan Sheng and
                  Yue Gu and
                  Chong Deng and
                  Wen Wang and
                  Shiliang Zhang and
                  Zhijie Yan and
                  Jingren Zhou},
  title        = {CosyVoice 2: Scalable Streaming Speech Synthesis with Large Language
                  Models},
  journal      = {CoRR},
  volume       = {abs/2412.10117},
  year         = {2024},
  url          = {https://doi.org/10.48550/arXiv.2412.10117},
  doi          = {10.48550/ARXIV.2412.10117},
  eprinttype    = {arXiv},
  eprint       = {2412.10117},
  bibsource    = {dblp computer science bibliography, https://dblp.org}
}

@inproceedings{AadmW,
  author       = {Ilya Loshchilov and
                  Frank Hutter},
  title        = {Decoupled Weight Decay Regularization},
  booktitle    = {7th International Conference on Learning Representations, {ICLR} 2019,
                  New Orleans, LA, USA, May 6-9, 2019},
  publisher    = {OpenReview.net},
  year         = {2019},
  url          = {https://openreview.net/forum?id=Bkg6RiCqY7},
  bibsource    = {dblp computer science bibliography, https://dblp.org}
}

@article{paddleocr,
  author       = {Cheng Cui and
                  Ting Sun and
                  Manhui Lin and
                  Tingquan Gao and
                  Yubo Zhang and
                  Jiaxuan Liu and
                  Xueqing Wang and
                  Zelun Zhang and
                  Changda Zhou and
                  Hongen Liu and
                  Yue Zhang and
                  Wenyu Lv and
                  Kui Huang and
                  Yichao Zhang and
                  Jing Zhang and
                  Jun Zhang and
                  Yi Liu and
                  Dianhai Yu and
                  Yanjun Ma},
  title        = {PaddleOCR 3.0 Technical Report},
  journal      = {CoRR},
  volume       = {abs/2507.05595},
  year         = {2025},
  url          = {https://doi.org/10.48550/arXiv.2507.05595},
  doi          = {10.48550/ARXIV.2507.05595},
  eprinttype    = {arXiv},
  eprint       = {2507.05595},
  bibsource    = {dblp computer science bibliography, https://dblp.org}
}

@article{qwen2,
  author       = {Team, Qwen and others},
  title        = {Qwen2 Technical Report},
  journal      = {CoRR},
  volume       = {abs/2407.10671},
  year         = {2024},
  url          = {https://doi.org/10.48550/arXiv.2407.10671},
  doi          = {10.48550/ARXIV.2407.10671},
  eprinttype    = {arXiv},
  eprint       = {2407.10671},
  bibsource    = {dblp computer science bibliography, https://dblp.org}
}

@article{hurst2024gpt,
  title={Gpt-4o system card},
  author={Hurst, Aaron and Lerer, Adam and Goucher, Adam P and Perelman, Adam and Ramesh, Aditya and Clark, Aidan and Ostrow, AJ and Welihinda, Akila and Hayes, Alan and Radford, Alec and others},
  journal={arXiv preprint arXiv:2410.21276},
  year={2024}
}

@article{MindWithEyes,
  author       = {Zhiyu Lin and
                  Yifei Gao and
                  Xian Zhao and
                  Yunfan Yang and
                  Jitao Sang},
  title        = {Mind with Eyes: from Language Reasoning to Multimodal Reasoning},
  journal      = {CoRR},
  volume       = {abs/2503.18071},
  year         = {2025},
  url          = {https://doi.org/10.48550/arXiv.2503.18071},
  doi          = {10.48550/ARXIV.2503.18071},
  eprinttype   = {arXiv},
  eprint       = {2503.18071},
  bibsource    = {dblp computer science bibliography, https://dblp.org}
}

\clearpage
\appendix
\section*{Appendix}

\section{Details of SlideASR-Bench}
\label{sec:app-prompts-for-slideasr-s}

\textbf{Statistics of SlideASR-Bench.} 
Table~\ref{tab:SlideASR-Bench-Detail} presents the key statistical metrics for SlideASR-Bench, including the number of samples, entities, and hours.
\begin{table}[h]\small
    \caption{Details of our proposed SlideASR-Bench.}
    \centering    \begin{tabular}{l|rrr}  
    \toprule
        \textbf{Subset}&  \textbf{Sample}& \textbf{Entity}& \textbf{Hour}\\ \midrule
         SlideASR-S (Training set) & 6,413 & 44,240 & 67.3 \\
         SlideASR-S (Test set) & 2,054 & 13,895 & 18.5 \\
         SlideASR-R & 60& 200 & 0.35\\

    \bottomrule
    \end{tabular}
    \label{tab:SlideASR-Bench-Detail}
\end{table}

\begin{figure}[h]
    \centering
    \fbox{\includegraphics[width=.9\linewidth]{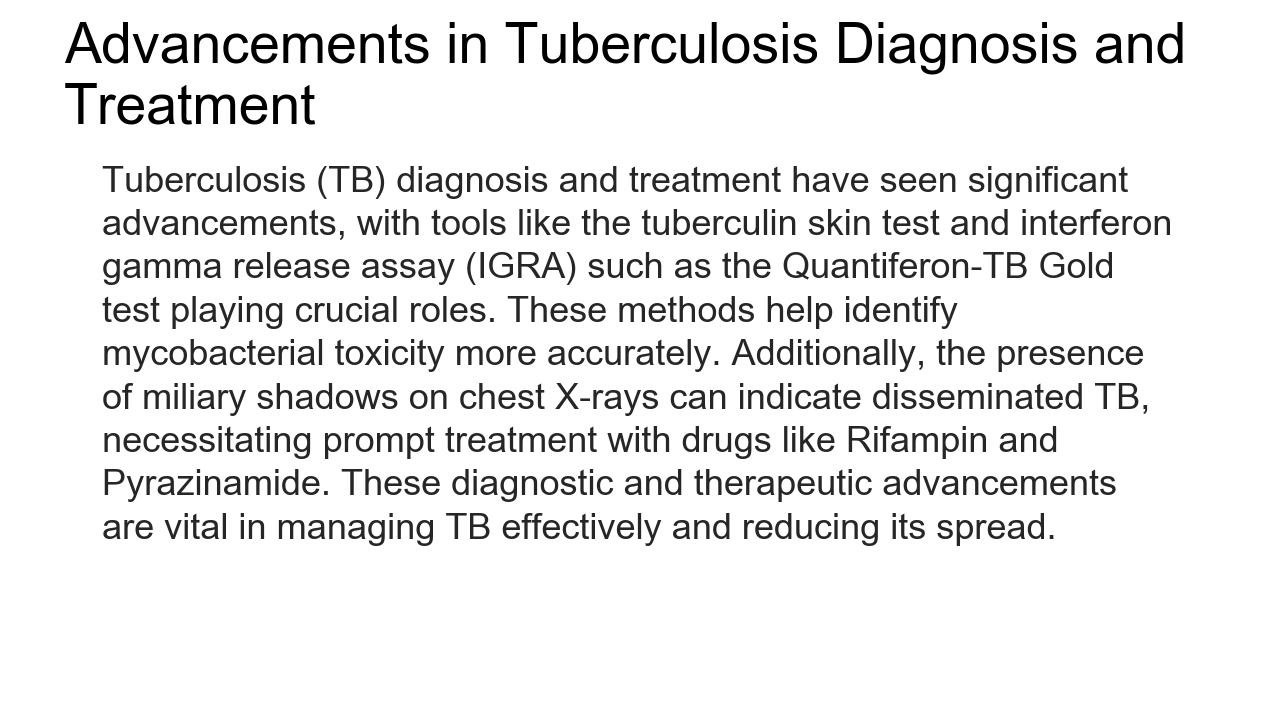}}
    \caption{Data example of SlideASR-S.}
    \label{fig:example-of-SlideASR-S}
\end{figure}

\noindent\textbf{Prompt for Generating SlideASR-S.}
The prompt for the LLM to generate text paragraphs based on a domain label and an entity list is as follows. Fig.~\ref{fig:example-of-SlideASR-S} shows an example of SlideASR-S.

\begin{figure}[h]
\begin{AIbox}{Prompt for Qwen2-14B-Instruct to generate slide text}
{Given a domain label and a list of entities, generate a title and a paragraph for use in a PPT report, with the requirement that the paragraph includes these entities, Keep paragraphs within 150 words.\\
Domain label:\\
\{\}\\
List of entities:\\
\{\}\\
Output format:\\
\#\#\#\\
Title\\
\#\#\#\\
Paragraph\\
}
\end{AIbox}
\end{figure}

\section{Evaluation Details}
\label{sec:app-prompts}
\subsection{Prompts}

The prompts for baseline models and our VAPO models are as follows.
\begin{figure}[h]
\begin{AIbox}{Prompt for baseline models}
\textbf{\#\#\# Contextless}\\
{Convert the audio to text.}\\

\textbf{\#\#\# Slide text as context}\\
{The speech is the speaker's talk accompanied by a slide, with the text of the slide being: \{\}\\
Transcribe the speech into text by integrating the speech with the slide content.}\\

\textbf{\#\#\# Slide image as context}\\
{Taking the image content into account, convert the audio to text.}
\end{AIbox}
% \caption{Template used to generate actions for the Re$^2$Search agent.}
\label{fig:prompt_research}
\end{figure}
\begin{figure}[h]
\begin{AIbox}{Prompt for VAPO model}
\textbf{\#\#\# Slide image as context}\\

\textbf{Role:System}\\
Your task is to convert the speech into text, and the image serves as the reference content related to the speech.\\
\textbf{Role:User}\\
First, recognize the text in the image and output it within <think> </think>. Then, referring to the thinking content, output the speech recognition result within <answer> </answer>.

\end{AIbox}
\label{fig:prompt_research}
\end{figure}

\subsection{Metrics}
\label{sec:app-metrics}
For SlideSpeech, as in the original work~\cite{slidespeech}, we use four metrics
\begin{itemize}
    \item \textbf{WER:} word error rate.
    \item \textbf{U-WER:} unbiased word error rate, computed on non-keyword segments, to evaluate model impact on general transcription.
    \item \textbf{B-WER:} unbiased word error rate, which measures errors on keyword spans.
    \item  \textbf{Recall:} keyword recall, the percentage of keywords fully and correctly recognized.
\end{itemize}
~\\
\noindent For SlideASR-Bench, we maintain consistency with ContextASR-Bench~\cite{ContextASR-Bench} and use the following three evaluation metrics:
\begin{itemize}
    \item \textbf{WER}: word error rate. For Chinese samples, we treat each character as a word.
    \item \textbf{NE-WER}: WER of named entity portion, we first perform a fuzzy match to identify key entities (with an edit distance tolerance of  $\frac{2}{WordCountOfEntity} - 1$) in the model's output, and then calculate the WER based on the fuzzy-matched entities.
    \item \textbf{NE-FNR}: The false negative ratio of named entities, calculated as $1 - \frac{H}{N}$, where $H$ and $N$ denote the recognized and ground-truth entity counts.
\end{itemize}

\subsection{Baselines}
\textbf{LALMs.} For LALMs, we selected Qwen2-Audio~\cite{qwen2audio} and Mi-Dasheng~\cite{MiDashengLM} as baselines, both with 7B parameters. ASR is a core capability of these models. Additionally, they have instruction-following abilities, making them suitable for the context-enhanced ASR task, i.e., \textit{Slide text as context} setting. For SlideSpeech, we additionally selected LCB-net~\cite{LCB-Net} and MaLa-ASR~\cite{MaLa-ASR} as baselines. These models were trained on the SlideSpeech training set, and the results for Dev and Test sets are provided~\cite{MaLa-ASR}.\\

\noindent\textbf{OLLMs.} For OLLMs, we selected MiniCPM-o-2.6~\cite{minicpmo}, Qwen2.5-Omni-3B~\cite{qwen25omni}, Qwen2.5-Omni-7B~\cite{qwen25omni} and Qwen3-Omni-30B-3B~\cite{qwen3omni} as baselines. Similarly, these models not only have ASR capabilities and instruction-following abilities, but they can also directly accept both image and audio as inputs.

\section{Results on ChineseLips}
\label{sec:app-chineselips}
Table~\ref{tab:app-ChineseLips} presents the results on ChineseLips~\cite{chineselips}. Similar to SlideSpeech~\cite{slidespeech}, ChineseLips is a real-world general-domain SlideASR dataset with low entity density both in the speech and the slides. Since ChineseLips does not provide text information for the slides, we report the CER metric on the transcribed text. 

The results show that our method achieves the lowest CER, demonstrating its effectiveness in real-world general-domain scenarios.

\begin{table}[!htb]\small
    \caption{Results on ChineseLips, a real-world Chinese SlideASR dataset. The best and second-best results are in \textbf{bold} and \ul{underlined}, respectively.}
    
    \centering
    
    \begin{tabular}{l c}
    \toprule
        \textbf{Model}& \textbf{CER}$\downarrow$\\\midrule
        \rowcolor{gray!13}\multicolumn{2}{c}{{\textbf{Contextless}}}\\
        Qwen2-Audio& 12.536 
\\
        Mi-Dasheng& 3.311 
\\
        MiniCPM-o-2.6& 2.252 
\\
        Qwen2.5-Omni-3B& 1.937 
\\
        Qwen2.5-Omni-7B& 2.243 
\\
        Qwen3-Omni-30B-A3B& 2.202 \\\midrule
        
        \rowcolor{gray!13}\multicolumn{2}{c}{\textit{\textbf{Slide text as context (Pipeline) }}}\\
        Qwen2-Audio & 84.291 
\\
        Mi-Dasheng & 65.505 \\
        Qwen3-Omni-30B-A3B& 69.172\\\midrule
        
        \rowcolor{gray!13}\multicolumn{2}{c}{\textit{\textbf{Slide image as context (End-to-End)}}}\\
        MiniCPM-o-2.6& 76.203 
\\
        Qwen2.5-Omni-3B& 24.847 
\\
        Qwen2.5-Omni-7B& 14.340 
\\
        Qwen3-Omni-30B-A3B& 41.930 
\\
        \textbf{VAPO-3B (Ours)}& \ul{1.548}\\
        \textbf{VAPO-7B (Ours)}& \bf1.298\\
    
    \bottomrule
    \end{tabular}
    
    \label{tab:app-ChineseLips}
\end{table}

\section{Case Study}
\label{sec:app-case-study}
\begin{figure*}[!h]
    \centering
    \includegraphics[width=1\linewidth]{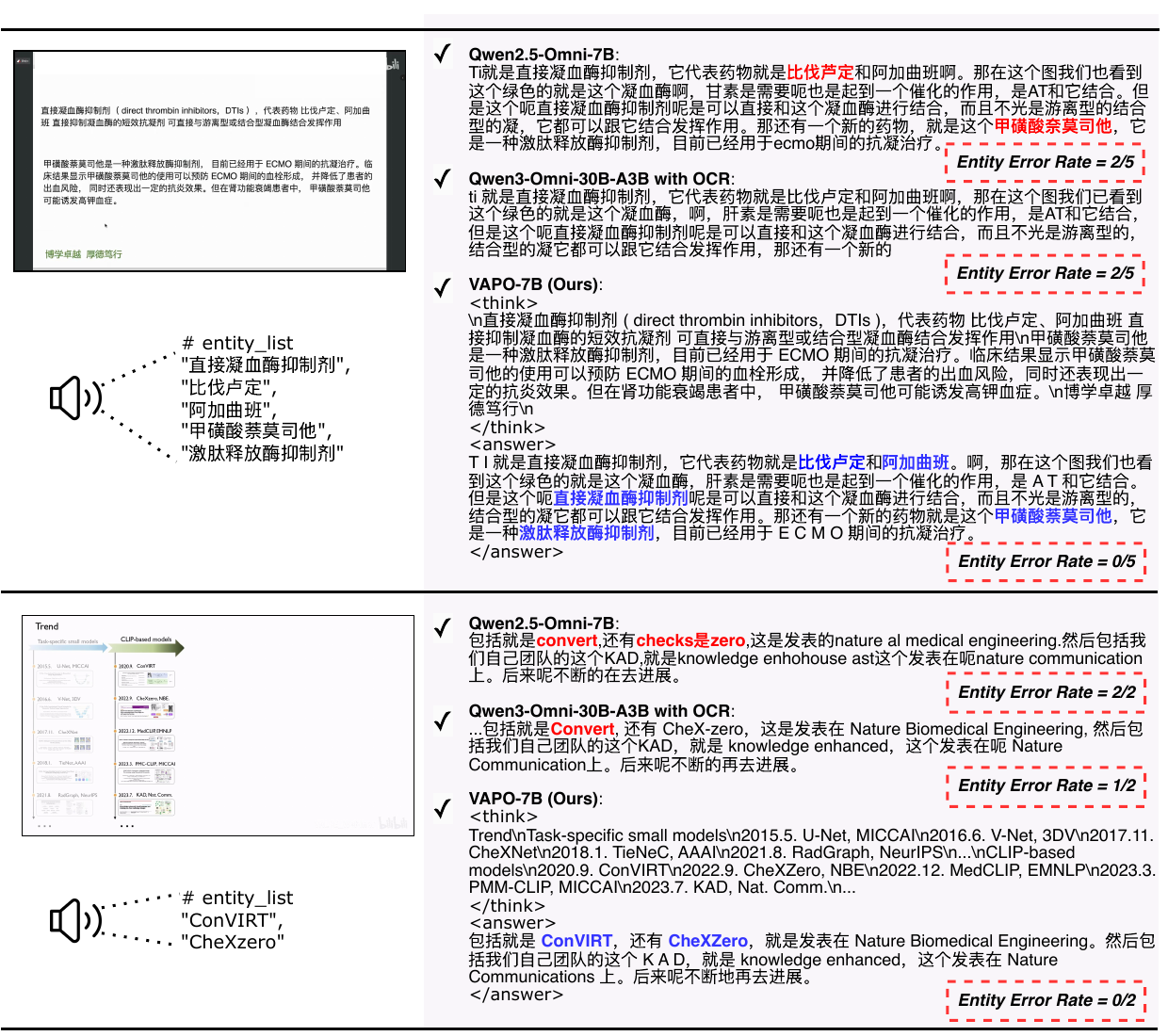}
    \caption{Qualitative comparison on the SlideASR-R dataset. We compare the transcription results of the baseline (Qwen2.5-Omni-7B), a strong contextual baseline (Qwen3-Omni-30B-A3B with OCR), and our VAPO-7B. \textbf{Red text} indicates entity recognition errors (e.g., misrecognizing "CheXzero" as "checks zero"), while \textbf{blue text} highlights correctly transcribed entities.}
    \label{fig:app-case-study}
\end{figure*}
\begin{figure*}[!h]

    \includegraphics[width=1.0\linewidth]{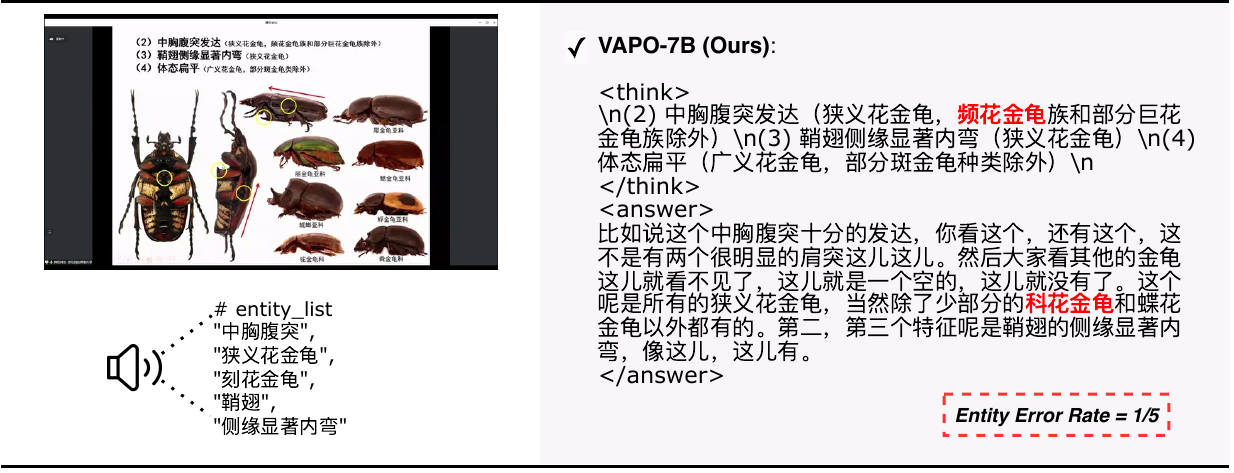}
    \caption{A failure case from SlideASR-R. OCR errors occurred due to low image resolution and small entity font size, leading to the loss of correct visual anchor points.}
    
    \label{fig:app-case-bad}
\end{figure*}

\subsection{Successful Case}

Fig.~\ref{fig:app-case-study} shows a comparison of outputs from Qwen2.5-Omni-7B, and Qwen3-Omni-30B-A3B and our proposed VAPO-7B models on samples from the SlideASR-R dataset. Among them, Qwen2.5-Omni-7B uses audio-only input, Qwen3-Omni-30B-A3B uses OCR text extracted from the slide image as context, and VAPO-7B uses the slide image as context input.  For Qwen2.5-Omni-7B, due to the lack of auxiliary information, the entity error rate is relatively high. For example, it misrecognized "ConVIRT" as "convert".  For Qwen3-Omni-30B-A3B, although slide text is used as context, it fails to utilize it effectively. For example, it also misrecognized "ConVIRT" as "Convert".  The VAPO-7B model achieves higher entity recognition accuracy thanks to its "Look before Transcription" reasoning structure.

\subsection{Failure Case}
Fig.~\ref{fig:app-case-bad} reveals a failure mode of our VAPO-7B model, originating from a visual perception error. The issue begins in the \textit{<think>} block, where the internal OCR component misidentifies a key entity. Specifically, the correct character (pronounced ke) is misrecognized as a different character (pronounced pin). This OCR error is highly plausible because the two characters are visually confusable due to their structural similarity. This type of resemblance is a known challenge for OCR systems, especially with low-resolution text, and it results in the absence of a correct visual anchor.

Due to the lack of a correct visual anchor in the \textit{<think>} block, the entity transcription in the \textit{<answer>} block ends up being incorrect. This case highlights that VAPO's performance is sensitive to low resolution and font size, particularly with visually similar characters. However, this doesn't diminish its overall advantage, as a contextless audio-only model would be equally, if not more, prone to failure when confronted with such inherent ambiguities in the source modalities.

\begin{table*}[t]\small
    \caption{Ablation results of functions on the SlideSpeech dataset.}
    \centering
    \begin{tabular}{ccc cccccccc}
    \toprule
    
 \textbf{ASR}& \textbf{OCR}& \textbf{VA}& \multicolumn{4}{c}{\textbf{Dev set}}& \multicolumn{4}{c}{\textbf{Test set}}\\
        \cmidrule(lr){4-7} \cmidrule(lr){8-11}
         \textbf{Reward}& \textbf{Reward}&\textbf{Reward}& \fontsize{8}{9}{\textbf{WER}}$\downarrow$& \fontsize{8}{9}{\textbf{B-WER}}$\downarrow$& \fontsize{8}{9}{\textbf{U-WER}}$\downarrow$&\fontsize{8}{9}{\textbf{Recall}}$\uparrow$& \fontsize{8}{9}{\textbf{WER}}$\downarrow$&\fontsize{8}{9}{\textbf{B-WER}}$\downarrow$& \fontsize{8}{9}{\textbf{U-WER}}$\downarrow$&\fontsize{8}{9}{\textbf{Recall}}$\uparrow$\\\midrule
         \ding{56}& \ding{56}& \ding{56}& 12.22 &12.74 & 5.26 &95.17 & 19.99 &20.71 & 9.80 &94.44 \\
         \ding{52}& \ding{56}& \ding{56}& 10.22 &10.68 & 4.01 &96.08 & 11.02 &11.52 & 3.92 &96.17 \\
         \ding{52}& \ding{52}& \ding{56}& 9.97 &10.49 & 3.83 &96.30 & 11.74 &12.25 & 4.47 &95.63 \\
         \ding{52}& \ding{52}& \ding{52}& \bf9.84&\bf10.31& \bf3.61&\bf96.54& \bf10.73&\bf11.24& \bf3.55&\bf96.57\\
 \bottomrule
    \end{tabular}
    \label{tab:app-ablation-slidespeech}
\end{table*}

\begin{table*}[t]\small
    \caption{Ablation results of the rewards on the SlideASR-S dataset.}
    \centering
    \begin{tabular}{ccc cccccc}
    \toprule
    
        \textbf{ASR}& \textbf{OCR}& \textbf{VA}& \multicolumn{3}{c}{\textbf{SlideASR-S (en)}}& \multicolumn{3}{c}{\textbf{SlideASR-S (zh)}}\\
        \cmidrule(lr){4-6} \cmidrule(lr){7-9}
         \textbf{Reward}& \textbf{Reward}&\textbf{Reward}& \fontsize{8}{9}{\textbf{WER}}$\downarrow$& \fontsize{8}{9}{\textbf{NE-WER}} $\downarrow$& \fontsize{8}{9}{\textbf{NE-FNR}}$\downarrow$& \fontsize{8}{9}{\textbf{WER}}$\downarrow$& \fontsize{8}{9}{\textbf{NE-WER}}$\downarrow$&\fontsize{8}{9}{\textbf{NE-FNR}}$\downarrow$\\\midrule
         \ding{56}& \ding{56}& \ding{56}& 100.08 &53.19 & 18.72 & 86.86 & 65.62 &9.62 \\
         \ding{52}& \ding{56}& \ding{56}& 5.40 &3.78 & 4.47 & 2.49 & 4.33 &2.49 \\
         \ding{52}& \ding{52}& \ding{56}& 4.98 &3.71 & 4.23 & 2.58 & 4.54 &2.82 \\
         \ding{52}& \ding{52}& \ding{52}& \bf4.90 &\bf3.19 & \bf3.73 & \bf2.47 & \bf4.21 &\bf2.22 \\
    \bottomrule
    \end{tabular}
    \label{tab:app-ablation-slideasr-s}
\end{table*}

\section{More Ablation Results on Reward Functions}
\label{sec:app-more-ablation}

Table~\ref{tab:app-ablation-slidespeech} and Table~\ref{tab:app-ablation-slideasr-s}  respectively present the ablation results of VAPO-3B (based on Qwen2.5-Omni-3B) on SlideSpeech~\cite{slidespeech} and SlideASR-S.  The results indicate that different reward functions have a positive impact on the final performance.

\section{More Cases of Attention Visualization}
\label{sec:app-more-attention-visualization}

\begin{figure*}[!thb]
    \centering
    \includegraphics[width=0.6\linewidth]{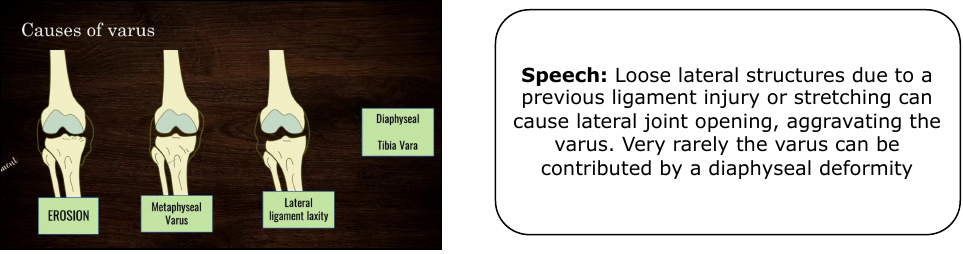}
    \vspace{15pt}
    \includegraphics[width=1\linewidth]{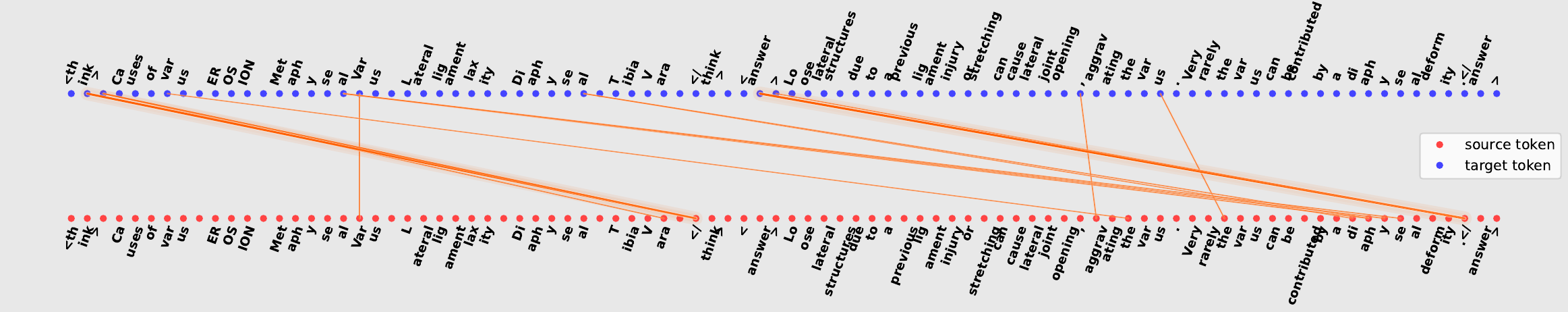}
    \includegraphics[width=0.6\linewidth]{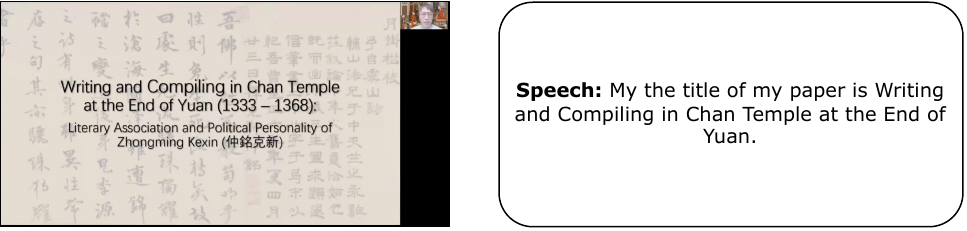}
    \includegraphics[width=1\linewidth]{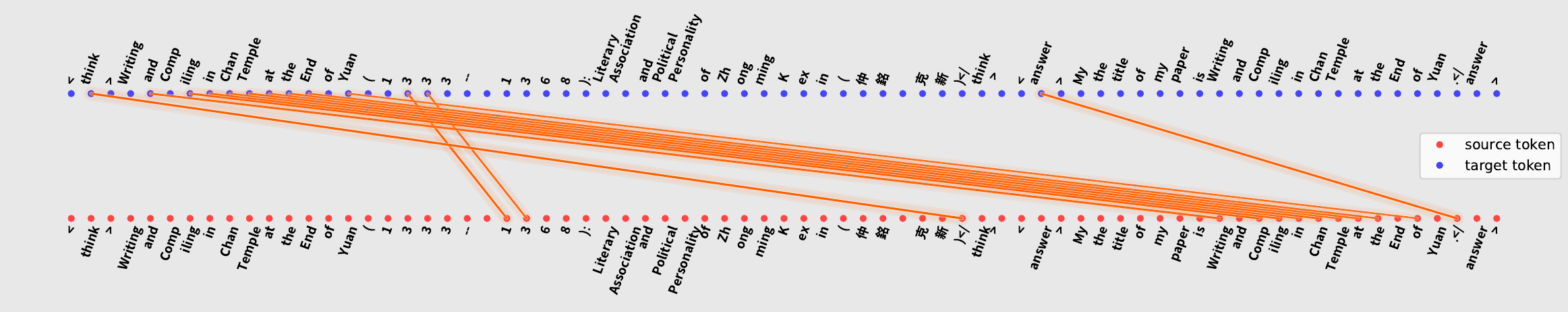}
    \caption{More cases of attention visualization.}
    \label{fig:app-attention-visualization}
\end{figure*}

Fig.~\ref{fig:app-attention-visualization} further presents two cases of attention visualization. It can be seen that when transcribing key entities, the model is able to focus its attention on the same entities in the \textit{<think>} block. This desirable property enables the model to accurately transcribe key entities in the speech.

\begin{table*}[!htb]\small
    \caption{Comparison of inference time and NE-FNR on the SlideASR-R dataset.}
    \centering
    \begin{tabular}{llcc}
    \toprule
    
          \textbf{Model}&\textbf{Setting}&  \textbf{\makecell{Inference time per sample (s)}}&\textbf{NE-FNR}\\ \midrule
        Qwen3-Omni-30B-A3B &Slide text as context& 105.98&28.22\\
 Qwen3-Omni-30B-A3B & Slide image as context& 172.95&24.75\\
        Qwen2.5-Omni-7B &Slide image as context&2.51&35.15\\
        VAPO-7B &Slide image as context& 7.27&15.35\\

    \bottomrule
    \end{tabular}
    \label{tab:app-latencey}
\end{table*}

\section{Inference Latency Analysis}
\label{sec:app-latency}

To evaluate the practical inference efficiency of our proposed VAPO framework, we measured the average inference time per sample on SlideASR-R and compared it against several key baseline models. The results are presented in Table~\ref{tab:app-latencey}.

As shown, our VAPO-7B model has an inference time of 7.27 seconds per sample. This is slower than Qwen2.5-Omni-7B (2.51s), which is expected, as the structured \textit{<think><answer>} generation process introduces a computational overhead. However, this moderate increase in latency is accompanied by a dramatic improvement in accuracy, with the NE-FNR dropping from 35.15 to 15.35.

More importantly, when compared to the best baseline model Qwen3-Omni-30B-A3B, our VAPO-7B is significantly more efficient and accurate. While not yet suitable for real-time applications, the latency of VAPO is a reasonable trade-off for its state-of-the-art performance, particularly for offline transcription tasks where accuracy is paramount.

\end{document}